\documentclass[aps,pra,reprint,showpacs,amsmath,amssymb,10pt]{revtex4-1}
\usepackage{graphicx}
\usepackage{bm}
\usepackage{slashed}
\usepackage{color}

\begin{document}
\title{Global phase and frequency comb structures in nonlinear Compton and Thomson scattering}
\author{K. Krajewska$^\dagger$}
\email[E-mail address:\;]{Katarzyna.Krajewska@fuw.edu.pl}
\author{M. Twardy$^\ddagger$}
\author{J. Z. Kami\'nski$^\dagger$}
\affiliation{$^\dagger$Institute of Theoretical Physics, Faculty of Physics, University of Warsaw, Ho\.{z}a 69,
00-681 Warszawa, Poland \\
$^\ddagger$Faculty of Electrical Engineering, Warsaw University of Technology,
Plac Politechniki 1, 00-661 Warszawa, Poland}
\date{\today}
\begin{abstract}
The Compton and Thomson radiation spectra, generated in collisions of an electron beam with a powerful laser beam, 
are studied in the framework of quantum and classical electrodynamics, respectively.
We show that there are frequency regimes where both radiation spectra are nearly identical, which for Compton scattering relates to the process
which preserves the electron spin. Although the radiation spectra are nearly identical, the corresponding probability amplitudes exhibit 
different global phases. This has pronounced consequences, which we demonstrate by investigating temporal power distributions in both cases.
We show that, contrary to Thomson scattering, it is not always possible to synthesize short laser pulses from Compton radiation.
This happens when the global phase of the Compton amplitude varies in a nonlinear way with the frequency 
of emitted photons. We also demonstrate that while the Compton process driven by a non-chirped laser pulse 
can generate chirped bursts of radiation, this is not the case for the Thomson process. In principle, both processes can 
lead to a generation of coherent frequency combs when single or multiple driving laser pulses
collide with electrons. Once we synthesize these combs into short bursts of radiation, we can control them, for instance, by changing
the time delay between the driving pulses.
\end{abstract}
\pacs{12.20.Ds, 12.90.+b, 42.55.Vc, 13.40.-f}
\maketitle

\section{Introduction}

Owing to the rapid development of high-power laser technology, in recent years we observe a renaissance of theoretical interest 
in studying strong-field quantum electrodynamics (QED) processes~\cite{fund1a,fund2,fund3}. Note that proceeding theoretical works were based
on the monochromatic plane wave approximation~\cite{Reiss,Ritus,BK,Kibble,Yakovlev}. However, with the parallel development of computational 
technology, it is possible now to extend these explorations and to study fundamental QED processes in multichromatic laser fields~\cite{KKphase,multi1} 
or in short laser pulses~\cite{puls1,puls2,puls3,KKcompton,KKpol,puls4}. New aspects of strong-field QED such as the electron (positron) 
polarization effects~\cite{KKpol,spin1}, the energy and angular correlations~\cite{corr1,corr2,KKangular,corr3,corr4}, the bremsstrahlung 
process at relativistically intense laser radiation~\cite{brem1}, or the electron-positron cascades~\cite{Ruhl2013}, are of great interest nowadays.
Usually, different types of cross sections or probability distributions are analyzed, leaving out problems related to the phase of probability 
amplitudes. However, in many cases, it is the global phase that plays a significant role. For instance, it is a very important parameter when studying coherence of 
high-order harmonics and the synthesis of attosecond pulses \cite{Farkas,KrauszIvanov,exp1,theo1}.

In this paper, we shall study the important role played by the global phase of probability amplitudes in nonlinear Compton 
scattering~\cite{KKcompton,KKpol} and its classical approximation -- nonlinear Thomson 
scattering~\cite{Sarachik1970,HK1996,HL1995,HF1995,SF1996,SF1997,SF2000,Umstadter2,Galkin2009,Chung2009,Lee2003,Lan2005,Kaplan2002,Liu2012}. 
For the classical theory we discuss the conditions of its applicability. We show that, in order to get comparable 
temporal power distributions from both theories, an extra condition on the Compton phase has to be imposed. 
We also propose the method of controlling the energy distribution of emitted radiation by properly modulated 
laser pulses such that it leads to the generation of coherent high-order harmonics combs. In addition, 
we investigate properties of the synthesized short radiation pulses.

The paper is organized as follows. In Sec.~\ref{pulse}, we define temporal and polarization properties 
of the laser pulses considered. Supplementary definitions of mutually orthogonal and normalized triad 
of vectors for two (in general elliptic) polarizations and for the direction of pulse propagation are discussed 
in Appendix~\ref{triads}. The basic theoretical scheme of the laser-induced QED Compton scattering is presented 
in Sec.~\ref{compton}, together with the derivation of the global phase for the Compton amplitude. We show 
that the total phase can be split into the kinematic and the dynamic parts. The kinematic phase is independent
of the electron spin degrees of freedom (hence, it is applicable also to the Compton scattering of spin-0 particles). While it can
be derived analytically (see, Eq.~\eqref{com5f}), the dynamic phase can be determined only numerically. 
For the pulses considered in this paper, the dynamic phase appears to be independent of the frequency of photons generated 
during the process. The reader can find more details in Appendix~\ref{diffcom}. The analogous analysis for nonlinear classical Thomson 
scattering is presented in Sec.~\ref{thomson} and Appendix~\ref{diffthom}. In Secs.~\ref{compton} and \ref{thomson}, 
we also compare the predictions of quantum and classical approaches for the energy distributions of emitted radiation, 
drawing particular attention to the different dependence of the global phase on the frequency of generated 
radiation. As we show, this is related to the quantum recoil of electrons during 
the emission of photons. We demonstrate in Sec.~\ref{synthesis} that the frequency dependence of the global phase plays a 
vital role in the temporal synthesis of generated radiation. We conclude that the quantum recoil effects result 
in a broader temporal distribution of radiation power for Compton scattering as compared to the predictions drawn 
from the classical theory. The interference of photons generated in Compton scattering by a modulated laser pulse 
(which consists of subpulses) is investigated in Sec.~\ref{combs}. In Sec.~\ref{combsdelay}, 
we show how the distance between the peaks in the energy spectrum can be controlled by the time delay of such subpulses. 
Although the presented results are for Compton scattering, we remark that a similar pattern is expected also for 
classical Thomson scattering provided that the frequencies are much smaller than the cut-off frequency for 
the quantum process. The analysis of the frequency combs in the laboratory frame is presented in Sec.~\ref{combslab}, 
with special emphasis on the partially angular-integrated energy distributions. This sort of investigation
is related to the fact that the frequency-comb structure is very sensitive to the direction of emission of 
generated radiation. Because quantum calculations are numerically demanding, in this analysis we choose 
frequencies much smaller than the cut-off frequency, which ensures that the classical calculations provide 
similar results. By doing this, we show that the comb structure survives the partial angular integration and, in principle, 
can be detected experimentally. Finally, in Sec.~\ref{conclusions} we draw some conclusions.

In analytical formulas we put $\hbar=1$. Hence, the fine-structure constant is $\alpha=e^2/(4\pi\varepsilon_0c)$. We use this constant 
also in expressions derived from classical electrodynamics, where it is meant to be multiplied by $\hbar$ when restoring the physical units.
Unless stated otherwise, in numerical analysis we use relativistic units (rel. units) such that $\hbar=m_{\rm e}=c=1$ where $m_{\rm e}$ is the electron mass.

\section{Laser pulse}
\label{pulse}

As in our previous investigations~\cite{KKphase,KKcompton,KKbw}, the laser pulse is described by the vector potential
\begin{equation}
\bm{A}(\phi)=A_0 B[\bm{\varepsilon}_1 f_1(\phi)+\bm{\varepsilon}_2 f_2(\phi)],
\label{las1}
\end{equation}
where the shape functions $f_j(\phi)$ vanish for $\phi<0$ and $\phi>2\pi$. The duration of the laser pulse, $T_{\mathrm{p}}$, introduces 
the fundamental frequency, $\omega=2\pi/T_{\mathrm{p}}$, such that
\begin{equation}
\phi=k\cdot x=\omega \Bigl( t-\frac{\bm{n}\cdot\bm{r}}{c}\Bigr),
\label{las2}
\end{equation}
in which the unit vector $\bm{n}$ points in the direction of propagation of the pulse. In a given reference frame, 
this direction is determined by the polar and azimuthal angles, $\theta_{\mathrm{L}}$ and $\varphi_{\mathrm{L}}$, respectively. 
This, according to Appendix \ref{triads}, settles the real polarization vectors $\bm{\varepsilon}_j=\bm{a}_j$ and $\bm{n}=\bm{a}_3$, Eq. \eqref{app1}. 
The constant $B>0$ is to be defined later. We also introduce the relativistically invariant parameter
\begin{equation}
\mu=\frac{|eA_0|}{m_{\mathrm{e}}c},
\label{las3}
\end{equation}
where $e=-|e|$ is the electron charge. With these notations, the electric and magnetic components of the laser pulse are equal to
\begin{equation}
\bm{\mathcal{E}}(\phi)=\frac{\omega m_{\mathrm{e}} c\mu}{e}B\bigl [\bm{\varepsilon}_1 f^{\prime}_1(\phi)+\bm{\varepsilon}_2 f^{\prime}_2(\phi)\bigr ],
\label{las4}
\end{equation}
and
\begin{equation}
\bm{\mathcal{B}}(\phi)=\frac{\omega m_{\mathrm{e}}\mu}{e}B\bigl [\bm{\varepsilon}_2 f^{\prime}_1(\phi)-\bm{\varepsilon}_1 f^{\prime}_2(\phi)\bigr ],
\label{las5}
\end{equation}
where '\textit{prime}' means the derivative with respect to $\phi$.

The shape functions are always normalized such that
\begin{equation}
\langle f_1^{\prime 2}\rangle +\langle f_2^{\prime 2}\rangle =\frac{1}{2},
\label{las6}
\end{equation}
where
\begin{equation}
\langle F \rangle=\frac{1}{2\pi}\int_0^{2\pi} F(\phi)\mathrm{d}\phi .
\label{las7}
\end{equation}
Note that, with such a normalization, the phase-averaged Poynting vector equals
\begin{equation}
\langle \bm{S}\rangle=\frac{1}{2}\varepsilon_0 c \Bigl(\frac{B\omega m_{\mathrm{e}}c\mu}{e}\Bigr)^2\bm{n}.
\label{las8}
\end{equation}

Laser pulses are also characterized by the number of oscillations of its electric or magnetic components, 
$N_{\mathrm{osc}}$. Together with the fundamental frequency $\omega$ (or, the pulse duration $T_{\mathrm{p}}$), 
they define the carrier frequency (or, the central frequency of a pulse), $\omega_{\mathrm{L}}=N_{\mathrm{osc}}\omega$. 
For a pulse generated by a given laser device, the carrier frequency is fixed whereas the remaining parameters can change. 
Therefore, it is useful to express the averaged intensity of the laser field, $I$, which is the modulus of the averaged Poynting 
vector, Eq.~\eqref{las8}, in terms of $\omega_{\mathrm{L}}$. Moreover, when comparing results for different shapes of laser 
pulses we have to impose extra conditions. For instance, by assuming that the flow of laser radiation per unit surface and unit 
time (i.e., the intensity $I$) for different durations of laser pulses is independent of $N_{\mathrm{osc}}$, we have to keep 
$B=N_{\mathrm{osc}}$. This leads to
\begin{equation}
I=\frac{m_{\mathrm{e}}^4c^6}{8\pi\alpha} \Bigl(\frac{\omega_{\mathrm{L}}}{m_{\mathrm{e}}c^2}\Bigr)^2 \mu^2 =A_{\mathrm{I}}\Bigl(\frac{\omega_{\mathrm{L}}}{m_{\mathrm{e}}c^2}\Bigr)^2 \mu^2,
\label{las9}
\end{equation}
with $A_{\mathrm{I}}\approx 2.3\times 10^{29}\mathrm{W/cm}^2$. On the other hand, if we assume that the averaged energy 
flow per unit surface (i.e., $W=T_{\mathrm{p}}I$) is fixed, we have to put $B=\sqrt{N_{\mathrm{osc}}}$, and
\begin{equation}
W=\frac{m_{\mathrm{e}}^3c^4}{4\alpha} \frac{\omega_{\mathrm{L}}}{m_{\mathrm{e}}c^2} \mu^2 =A_{\mathrm{W}} \frac{\omega_{\mathrm{L}}}{m_{\mathrm{e}}c^2} \mu^2,
\label{las10}
\end{equation}
with $A_{\mathrm{W}}\approx 1.9\times 10^{9}\mathrm{J/cm}^2$. The latter situation could be met in experiments,
as the energy of a laser pulse and a size of the laser focus are quite often given whereas the time-duration is changed.

In our numerical illustrations, we shall choose the linearly polarized laser pulse, $f_1(\phi)=f(\phi)$ and $f_2(\phi)=0$, and use the shape functions with the $\sin^2$-type envelope,
\begin{equation}
f'(\phi)\propto \sin^2\Bigl(N_{\mathrm{rep}}\frac{\phi}{2}\Bigr)\sin(N_{\mathrm{rep}}N_{\mathrm{osc}}\phi+\chi),
\label{las11}
\end{equation}
where the proportionality constant is determined by the normalization condition~\eqref{las6}. 
Here, $\chi$ denotes the carrier-envelope phase, $N_{\mathrm{rep}}$ determines the number of modulations in the pulse (or, the number of subpulses), 
and $N_{\mathrm{osc}}$ sets the number of cycles within the subpulse. $N_{\rm osc}$ also establishes the central frequency $\omega_{\mathrm{L}}=N_{\mathrm{osc}}\omega$, 
which is considered to be fixed and equal to $\omega_{\mathrm{L}}=1.548\mathrm{eV}$ in the laboratory frame. This corresponds to a Ti-Sapphire laser beam 
of wavelength $\lambda_{\mathrm{L}}=800\mathrm{nm}$. Let us also remark that, while for $\phi=0$ we have $f(0)=0$, 
for $\phi=2\pi$ the vector potential has to vanish as well. This is automatically satisfied if $N_{\mathrm{osc}}\geqslant 2$, 
whereas for $N_{\mathrm{osc}}=1$ the envelope phase $\chi$ must be equal to 0 or $\pi$. We also assume that $B=N_{\mathrm{rep}}N_{\mathrm{osc}}$, 
so that the averaged laser field intensity is independent of integers $N_{\mathrm{rep}}$ and $N_{\mathrm{osc}}$.

\section{Compton scattering}
\label{compton}

When scattering a laser pulse off a free electron, a nonlaser photon is detected. It is described by the wave four-vector $K$ and, in general, the elliptic polarization four-vector 
$\varepsilon_{{\bm K}\sigma}$ ($\sigma=1,2$) such that 
 \begin{equation}
K\cdot\varepsilon_{\bm{K}\sigma}=0,\quad \varepsilon_{\bm{K}\sigma}\cdot\varepsilon_{\bm{K}\sigma'}^*=-\delta_{\sigma\sigma'}.
\end{equation}
The wave four-vector $K$ satisfies the on-shell mass relation $K\cdot K=0$ as well as it defines the photon frequency $\omega_{\bm K}=cK^0=c|{\bm K}|$.
As shown in Ref.~\cite{KKbw}, $\varepsilon_{{\bm K}\sigma}$ can be chosen as the space-like vector, i.e., $\varepsilon_{{\bm K}\sigma}=(0,{\bm\varepsilon}_{{\bm K}\sigma})$.
The scattering is accompanied by the electron transition from the initial $({\rm i})$ to the final $({\rm f})$ state, each characterized by the four-momentum 
and the spin projection; $(p_{\rm i},\lambda_{\rm i})$ and $(p_{\rm f},\lambda_{\rm f})$. While moving in a laser pulse, the electron acquires additional 
momentum shift~\cite{KKcompton} (see, also Ref.~\cite{KKbw}) which leads to a notion of the laser-dressed momentum:
\begin{align}
\bar{p}=p - & \mu m_\mathrm{e} c\Bigl(\frac{p\cdot\varepsilon_1}{p\cdot k}\langle f_1\rangle 
+ \frac{p\cdot\varepsilon_2}{p\cdot k}\langle f_2\rangle\Bigr)k \nonumber \\ 
+ & \frac{1}{2}(\mu m_\mathrm{e} c)^2\frac{\langle f_1^2\rangle+\langle f_2^2\rangle}{p\cdot k}k .  \label{com0}
\end{align}
It was discussed in Refs.~\cite{KKbw,KKcompton} that the laser-dressed momenta~\eqref{com0} are gauge-dependent, and therefore they do not have clear physical meaning. 
Nevertheless, all formulas derived in Ref.~\cite{KKcompton} depend on the quantity 
\begin{equation}
 P_N=\bar{p}_{\rm i}-\bar{p}_{\rm f}+Nk-K,
\label{com4}
\end{equation}
where the difference $\bar{p}_{\rm i}-\bar{p}_{\rm f}$ is already gauge-invariant.

We take from our previous paper~\cite{KKcompton} the derivation of the Compton photon spectra. Hence, 
the frequency-angular distribution of energy of scattered photons for an unpolarized electron beam is given by the formula
\begin{equation}
\frac{\mathrm{d}^3E_{\mathrm{C}}}{\mathrm{d}\omega_{\bm{K}}\mathrm{d}^2\Omega_{\bm{K}}}=
\frac{1}{2}\sum_{\sigma=1,2}\sum_{\lambda_{\rm i}=\pm}\sum_{\lambda_{\rm f}=\pm}\frac{\mathrm{d}^3E_{\mathrm{C},\sigma}(\lambda_{\rm i},\lambda_{\rm f})}{\mathrm{d}\omega_{\bm{K}}\mathrm{d}^2\Omega_{\bm{K}}},
\label{com1}
\end{equation}
where
\begin{equation}
 \frac{\mathrm{d}^3E_{\mathrm{C},\sigma}(\lambda_{\rm i},\lambda_{\rm f})}{\mathrm{d}\omega_{\bm{K}}\mathrm{d}^2\Omega_{\bm{K}}}=\alpha\bigl|{\cal{A}}_{{\rm C},\sigma}(\lambda_{\rm i},\lambda_{\rm f})\bigr|^2,
\label{com2}
\end{equation}
and the scattering amplitude equals
\begin{equation}
 {\cal{A}}_{{\rm C},\sigma}(\lambda_{\rm i},\lambda_{\rm f})=\frac{m_{\rm e}cK^0}{\sqrt{p_{\rm i}^0k^0(k\cdot p_{\rm f})}}\mathfrak{A}_{\mathrm{C},\sigma}(\lambda_{\rm i},\lambda_{\rm f}) ,
\label{com3}
\end{equation}
with
\begin{equation}
\mathfrak{A}_{\mathrm{C},\sigma}(\lambda_{\rm i},\lambda_{\rm f})=\sum_N D_N\frac{1-{\rm e}^{-2\pi{\rm i}P_N^0/k^0}}{2\pi{\rm i}P_N^0/k^0}.
\label{com3a}
\end{equation}
The scattering amplitude is expressed as a Fourier series; for the coefficients $D_N$, the reader is referred to Eqs. (23) and (44) in Ref.~\cite{KKcompton}.
$P_N^0$ is obtained from Eq.~\eqref{com4}.

\begin{figure}
\includegraphics[width=8cm]{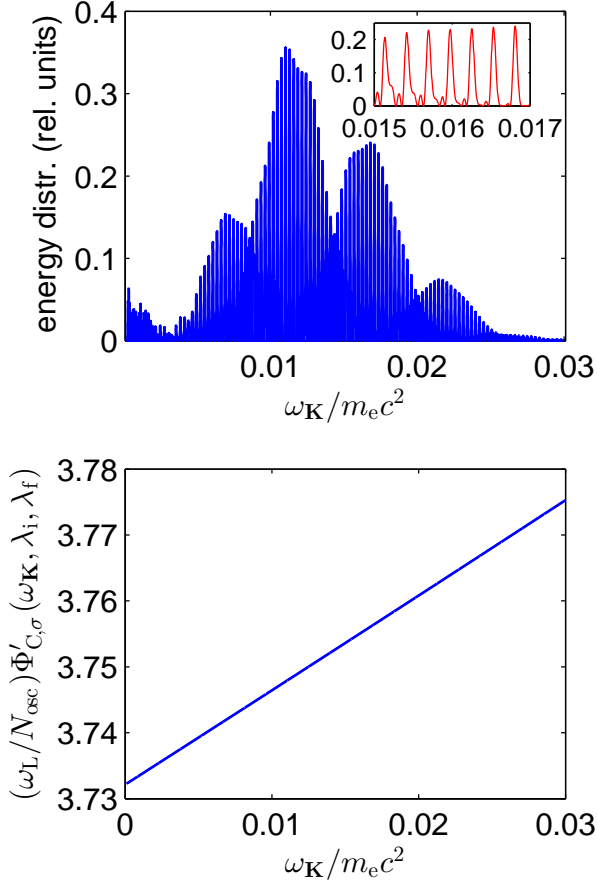}%
\caption{(Color online) The energy distribution, Eq.~\eqref{com2}, for the non-spin flipping Compton process
(upper panel; in the inset an enlarged portion of the distribution is presented), and the derivative of the phase 
as a function of the frequency $\omega_{\bm{K}}$ (lower panel), Eq.~\eqref{com5}. The laser pulse is linearly polarized 
in the $x$-direction ($\bm{\varepsilon}_1=\bm{e}_x$) and it propagates in the $z$-direction. The electron beam propagates 
in the opposite direction. The calculation is performed in the reference frame of electrons for the laser pulse carrier 
frequency $\omega_{\mathrm{L}}=4.15\times 10^{-4}m_{\mathrm{e}}c^2$, and for $\mu=2$, $N_{\mathrm{osc}}=16$, $N_{\mathrm{rep}}=1$, and
$\chi=0$. The scattered Compton photon is emitted in the direction $\theta_{\bm{K}}=0.2\pi$ and $\varphi_{\bm{K}}=0$. 
In the laboratory frame, for the Ti-Sapphire laser pulse of the central frequency 1.548eV, these parameters correspond 
to the electron beam's energy 35MeV, whereas the Compton photon of frequency $0.02m_{\mathrm{e}}c^2$ relates to 
$\omega^{\mathrm{LAB}}_{\bm{K}}=0.134\mathrm{MeV}$ and $\theta^{\mathrm{LAB}}_{\bm{K}}=0.986\pi$. In the electron beam 
reference frame, the cut-off frequency is $\omega_{\mathrm{cut}}\approx 5.24m_{\mathrm{e}}c^2$.
\label{qspecf2comb20130914}}
\end{figure}

\begin{figure}
\includegraphics[width=8cm]{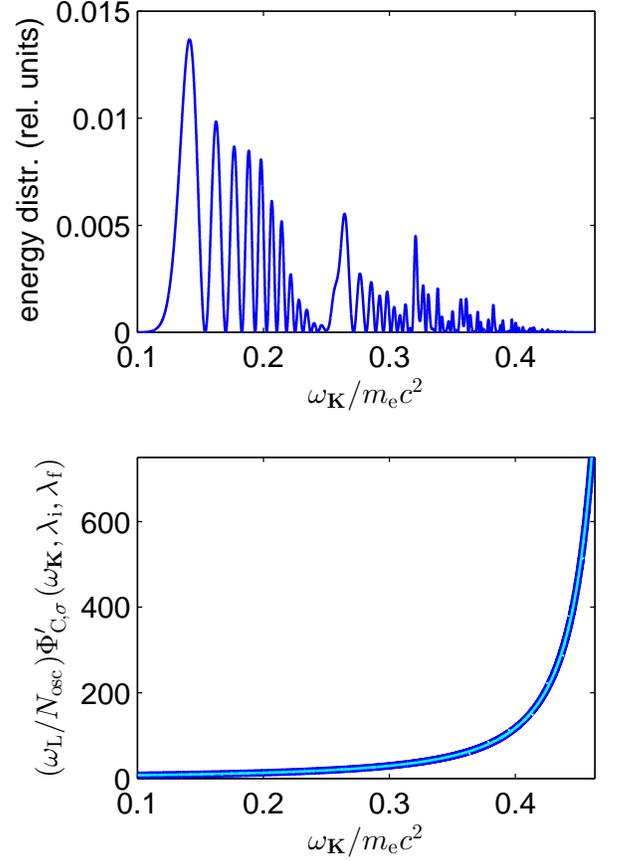}%
\caption{(Color online) The same as in Fig.~\ref{qspecf2comb20130914}, but for $\omega_{\mathrm{L}}=4.15\times 10^{-1}m_{\mathrm{e}}c^2$, $\mu=1$, and  
$\theta_{\bm{K}}=0.99\pi$. In the laboratory frame, for the Ti-Sapphire laser pulse of the central frequency 1.548eV, these parameters correspond 
to the electron beam's energy 35GeV. In the lower panel, the thick blue (dark) line represents the derivative of the Compton phase, Eq.~\eqref{com5}, 
on which the thin cyan (gray) line is overprinted, representing the kinematic Compton phase, Eq.~\eqref{com5e}. In the electron frame, 
the cut-off frequency is $\omega_{\mathrm{cut}}\approx 0.5m_{\mathrm{e}}c^2$.
\label{qspecf1xcomb20130914}}
\end{figure}

Eq.~\eqref{com3} allows one to define the phase $-\pi < \Phi_{\mathrm{C},\sigma} \leqslant\pi$ of the Compton scattering amplitude,
\begin{align}
\Phi_{\mathrm{C},\sigma}(\omega_{\bm{K}},\lambda_{\rm i},\lambda_{\rm f})= & \arg (\mathcal{A}_{\mathrm{C},\sigma}(\lambda_{\rm i},\lambda_{\rm f})) \nonumber \\
 = & \arg (\mathfrak{A}_{\mathrm{C},\sigma}(\lambda_{\rm i},\lambda_{\rm f})),
\label{com5}
\end{align}
which depends on the electron spin degrees of freedom. This phase is gauge and relativistically invariant, and can be split into two parts, 
if we rewrite $\mathfrak{A}_{\mathrm{C},\sigma}(\lambda_{\rm i},\lambda_{\rm f})$ as
\begin{equation}
\mathfrak{A}_{\mathrm{C},\sigma}(\lambda_{\rm i},\lambda_{\rm f})=\mathrm{e}^{\mathrm{i}\pi N_{\mathrm{eff}}}\sum_{N} (-1)^N D_N \mathrm{sinc}[\pi (N-N_{\mathrm{eff}})],
\label{com5a}
\end{equation}
where $\mathrm{sinc}(x)=\sin (x)/x$, and
\begin{equation}
N_{\mathrm{eff}}=(K^0+\bar{p}^0_{\mathrm{f}}-\bar{p}^0_{\mathrm{i}})/k^0.
\label{com5b}
\end{equation}
With this factorization, the Compton phase becomes
\begin{equation}
\Phi_{\mathrm{C},\sigma}(\omega_{\bm{K}},\lambda_{\rm i},\lambda_{\rm f})
=\Phi^{\mathrm{kin}}_{\mathrm{C}}(\omega_{\bm{K}}) + \Phi^{\mathrm{dyn}}_{\mathrm{C},\sigma}(\omega_{\bm{K}},\lambda_{\rm i},\lambda_{\rm f}),
\label{com5c}
\end{equation}
where
\begin{equation}
\Phi^{\mathrm{kin}}_{\mathrm{C}}(\omega_{\bm{K}})=\arg\bigl(\mathrm{e}^{\mathrm{i}\pi N_{\mathrm{eff}}} \bigr)=\pi N_{\mathrm{eff}}\ (\mathrm{mod}\ 2\pi)
\label{com5e}
\end{equation}
and
\begin{equation}
\Phi^{\mathrm{dyn}}_{\mathrm{C},\sigma}(\omega_{\bm{K}},\lambda_{\rm i},\lambda_{\rm f})=\arg\Bigl\{
\sum_{N} (-1)^N D_N \mathrm{sinc}[\pi (N-N_{\mathrm{eff}})]\Bigr\}.
\label{com5d}
\end{equation}
The former phase we call {\it kinematic}, as it depends only on the kinematics of the process and it is independent of the spin degrees of freedom 
(it remains the same for spin-0 particles). The latter phase we call {\it dynamic}. In general, the dynamic phase can be determined only numerically.
However, for linearly polarized laser pulses and linearly polarized emitted photons (both considered in this paper) this phase is frequency-independent.

On the other hand, by applying the momenta conservation laws (i.e., $k\cdot P_N=0$ and $\varepsilon_j\cdot P_N=0$ for $j=1,2$), one can show that
\begin{equation}
\Phi^{\mathrm{kin}}_{\mathrm{C}}(\omega_{\bm{K}})=\mathcal{F}\; \frac{\omega^{\mathrm{Th}}_{\bm{K}}}{\omega} \; (\mathrm{mod}\; 2\pi),
\label{com5f}
\end{equation}
where the Thomson (i.e., classical) frequency $\omega^{\mathrm{Th}}_{\bm{K}}$ \cite{KKscale} is equal to
\begin{equation}
\omega^{\mathrm{Th}}_{\bm{K}}=\frac{N_{\mathrm{eff}}ck\cdot p_{\mathrm{i}}}
{n_{\bm{K}}\cdot [\bar{p}_{\mathrm{i}}+\mu m_{\mathrm{e}}c(\langle f_1\rangle\varepsilon_1+\langle f_2\rangle\varepsilon_2)]},
\label{com5gg}
\end{equation}
or
\begin{equation}
\omega^{\mathrm{Th}}_{\bm{K}}=\frac{\omega_{\bm{K}}}{1-\omega_{\bm{K}}/\omega_{\mathrm{cut}}},
\label{com5g}
\end{equation}
where
\begin{equation}
\omega_{\mathrm{cut}}=\frac{cp_{\mathrm{i}}\cdot n}{n_{\bm{K}}\cdot n}
\label{com5h}
\end{equation}
is the \textit{cut-off} frequency for the Compton spectra, $\omega_{\bm{K}}<\omega_{\mathrm{cut}}$ \cite{KKscale}. In Eq.~\eqref{com5f}, $\mathcal{F}$ is independent of $\omega_{\bm{K}}$,
\begin{equation}
\mathcal{F}=\pi\ \frac{p_{\mathrm{i}}\cdot n_{\bm{K}}}{p_{\mathrm{i}}\cdot n} (1+\mathcal{F}_1+\mathcal{F}_2+\mathcal{F}_{\mathrm{sq}}),
\label{com5i}
\end{equation}
with
\begin{equation}
\mathcal{F}_j=\mu m_{\mathrm{e}}c\langle f_j\rangle \Bigl(\frac{n_{\bm{K}}\cdot\varepsilon_j}{n_{\bm{K}}\cdot p_{\mathrm{i}}}- 
\frac{p_{\mathrm{i}}\cdot\varepsilon_j}{p_{\mathrm{i}}\cdot n}
\frac{n_{\bm{K}}\cdot n}{n_{\bm{K}}\cdot p_{\mathrm{i}}}\Bigr),\ j=1,2\ ,
\label{com5j}
\end{equation}
and
\begin{equation}
\mathcal{F}_{\mathrm{sq}}=\frac{1}{2}(\mu m_{\mathrm{e}}c)^2(\langle f^2_1\rangle + \langle f^2_2\rangle )
\frac{1}{p_{\mathrm{i}}\cdot n}\frac{n_{\bm{K}}\cdot n}{n_{\bm{K}}\cdot p_{\mathrm{i}}} .
\label{com5k}
\end{equation}

In Figs.~\ref{qspecf2comb20130914} and~\ref{qspecf1xcomb20130914}, we show the results for the spin-conserved ($\lambda_{\mathrm{i}}\lambda_{\mathrm{f}}=1$) 
Compton process in the electron beam reference frame. In Fig.~\ref{qspecf2comb20130914}, we choose the frequency range much smaller than the cut-off frequency, 
Eq.~\eqref{com5h}. One can see that, for this range of frequencies, the derivative of the Compton phase linearly depends on the emitted photon frequency,
$\omega_{\bm{K}}$. Since $\omega_{\bm{K}}\ll\omega_{\mathrm{cut}}$, one can expect that the classical theory will give almost an identical result. 
On the other hand, Fig. \ref{qspecf1xcomb20130914} presents the results for frequencies comparable to the cut-off value. As we see, when $\omega_{\bm{K}}$ is 
approaching $\omega_{\mathrm{cut}}$, the derivative of the Compton phase (and the phase itself) starts to depend nonlinearly on the Compton photon 
frequency and tends to infinity when $\omega_{\bm{K}}\rightarrow\omega_{\mathrm{cut}}$. Moreover, as mentioned above, for the considered linear polarizations 
of the laser and scattered radiation, the Compton phase, up to a constant term (i.e., independent of $\omega_{\bm{K}}$), is equal to the kinematic one.

\section{Thomson scattering}
\label{thomson}

As one can check in Ref.~\cite{LL2}, the acceleration $\bm{a}$ of an electron in arbitrary electric and magnetic fields, $\bm{\mathcal{E}}$ and $\bm{\mathcal{B}}$, is given by the formula
\begin{equation}
\bm{a}=\frac{e}{m_{\mathrm{e}}}\sqrt{1-\bm{\beta}^2}\bigl[\bm{\mathcal{E}}-\bm{\beta}(\bm{\beta}\cdot\bm{\mathcal{E}}) +c\bm{\beta}\times\bm{\mathcal{B}}\bigr].  
\label{thom1}
\end{equation}
Hence, the relativistic Newton-Lorentz equations which determine the classical trajectory of accelerated electrons can be rewritten in the form
\begin{align}
\frac{\mathrm{d}t(\phi)}{\mathrm{d}\phi}=&\frac{1}{\omega(1-\bm{n}\cdot\bm{\beta}(\phi))}, \label{thom2} \\
\frac{\mathrm{d}\bm{r}(\phi)}{\mathrm{d}\phi}=&\frac{c}{\omega}\frac{\bm{\beta}(\phi)}{1-\bm{n}\cdot\bm{\beta}(\phi)}, \nonumber \\
\frac{\mathrm{d}\bm{\beta}(\phi)}{\mathrm{d}\phi}=&\mu\frac{\sqrt{1-\bm{\beta}^2(\phi)}}{1-\bm{n}\cdot\bm{\beta}(\phi)} \nonumber \\
\times & \Bigl[ 
\bigl(\bm{\varepsilon}_1-\bm{\beta}(\phi)(\bm{\beta}(\phi)\cdot\bm{\varepsilon}_1)+\bm{\beta}(\phi)\times\bm{\varepsilon}_2\bigr)f^{\prime}_1(\phi) \nonumber \\
 +& \bigl(\bm{\varepsilon}_2-\bm{\beta}(\phi)(\bm{\beta}(\phi)\cdot\bm{\varepsilon}_2)-\bm{\beta}(\phi)\times\bm{\varepsilon}_1\bigr)f^{\prime}_2(\phi)\Bigr] .\nonumber
\end{align}
Here, the phase, $\phi=\omega(t-\bm{n}\cdot\bm{r}(t)/c)$, is used as an independent variable, instead of time $t$. 
The frequency-angular distribution of emitted radiation of polarization $\bm{\varepsilon}_{\bm{K},\sigma}$ is given by the Thomson 
formula~\cite{Jackson1975} (we use the same notation for the radiation emitted during this process as for the Compton scattering)
\begin{equation}
\frac{\mathrm{d}^3E_{\mathrm{Th},\sigma}}{\mathrm{d}\omega_{\bm{K}}\mathrm{d}^2\Omega_{\bm{K}}}=
\alpha\bigl | \mathcal{A}_{\mathrm{Th},\sigma} \bigr |^2 ,
\label{thom3}
\end{equation}
where
\begin{align}
\mathcal{A}_{\mathrm{Th},\sigma}=\frac{1}{2\pi}\int\limits_{0}^{2\pi}
\Upsilon_{\sigma}(\phi) \exp\Bigl[\mathrm{i}\omega_{\bm{K}}\frac{\ell(\phi)}{c}\Bigr] \mathrm{d}\phi ,
\label{thom4}
\end{align}
with
\begin{equation}
\Upsilon_{\sigma}(\phi)=\bm{\varepsilon}_{\bm{K},\sigma}^*\cdot\frac{\bm{n}_{\bm{K}}\times [(\bm{n}_{\bm{K}}-\bm{\beta}(\phi))\times \bm{\beta}^{\prime}(\phi)]}{\bigl(1-\bm{n}_{\bm{K}}\cdot \bm{\beta}(\phi)\bigr)^2} ,
\label{thom5}
\end{equation}
and
\begin{equation}
\ell(\phi)=ct(\phi) -\bm{n}_{\bm{K}}\cdot \bm{r}(\phi) .
\label{thom5a}
\end{equation}
Here, '\textit{prime}' means again the derivative with respect to the phase $\phi$.

Let us further define the position four-vector
\begin{equation}
x(\phi)=(ct(\phi),\bm{r}(\phi)).
\label{thom6}
\end{equation}
After some algebraic manipulations, we show that
\begin{equation}
\Upsilon_{\sigma}(\phi)=K^0\frac{(\varepsilon_{\bm{K},\sigma}\cdot x^{\prime\prime})(K\cdot x^{\prime}) - (\varepsilon_{\bm{K},\sigma}\cdot x^{\prime})(K\cdot x^{\prime\prime})}{(K\cdot x^{\prime})^2}.
\label{thom7}
\end{equation}
Now, we can present the Thomson formula in a manifestly relativistic form. To do so, we define the relativistically invariant quantities: 
$\Upsilon^{\mathrm{inv}}_{\sigma}(\phi)=\Upsilon_{\sigma}(\phi)/K^0$ and
\begin{equation}
\mathcal{A}^{\mathrm{inv}}_{\mathrm{Th},\sigma}=\frac{1}{2\pi}\int_{0}^{2\pi}
\Upsilon^{\mathrm{inv}}_{\sigma}(\phi) \mathrm{e}^{\mathrm{i}K\cdot x(\phi)} \mathrm{d}\phi .
\label{thom8}
\end{equation}
Hence, the frequency-angular distribution of radiated energy equals
\begin{equation}
\frac{\mathrm{d}^3E_{\mathrm{Th},\sigma}}{\mathrm{d}\omega_{\bm{K}}\mathrm{d}^2\Omega_{\bm{K}}}=
\alpha\frac{\omega_{\bm{K}}^2}{c^2}\bigl | \mathcal{A}^{\mathrm{inv}}_{\mathrm{Th},\sigma} \bigr |^2 .
\label{thom9}
\end{equation}
The advantage of the above formulation is that the invariant amplitude $\mathcal{A}^{\mathrm{inv}}_{\mathrm{Th},\sigma}$ 
can be calculated in the most convenient reference frame (for instance, in the reference frame of initial electrons), and afterwards 
transformed to another reference frame. It also leads to the simplifications for the invariant amplitude. Indeed, integrating by parts, we get
\begin{align}
\mathcal{A}^{\mathrm{inv}}_{\mathrm{Th},\sigma}=\frac{1}{2\pi} & \Bigl[
\frac{\varepsilon_{\bm{K},\sigma}\cdot x^{\prime}}{K\cdot x^{\prime}}\mathrm{e}^{\mathrm{i}K\cdot x}\Big|_0^{2\pi} \nonumber \\
 & -\mathrm{i}\int_{0}^{2\pi}
(\varepsilon_{\bm{K},\sigma}\cdot x^{\prime}) \mathrm{e}^{\mathrm{i}K\cdot x} \mathrm{d}\phi
\Bigr] ,
\label{thom10}
\end{align}
or, in a particular reference frame,
\begin{eqnarray}
\mathcal{A}^{\mathrm{inv}}_{\mathrm{Th},\sigma}&=&\frac{1}{2\pi}\frac{c}{\omega_{\bm{K}}}  \Bigl[
 -\frac{\bm{\varepsilon}_{\bm{K},\sigma}\cdot \bm{\beta}}{1-\bm{n}_{\bm{K}}\cdot\bm{\beta}}\mathrm{e}^{\mathrm{i}\omega_{\bm{K}}(t-\bm{n}_{\bm{K}}\cdot\bm{r}/c)}\Big|_0^{2\pi} \nonumber \\
 & +&\mathrm{i}\frac{\omega_{\bm{K}}}{\omega}\int_{0}^{2\pi}
\frac{\bm{\varepsilon}_{\bm{K},\sigma}\cdot \bm{\beta}}{1-\bm{n}\cdot\bm{\beta}} \mathrm{e}^{\mathrm{i}\omega_{\bm{K}}(t-\bm{n}_{\bm{K}}\cdot\bm{r}/c)} \mathrm{d}\phi
\Bigr] .
\label{thom11}
\end{eqnarray}
This is an analogue of Jackson's formula (Ref~\cite{Jackson1975}, Eq.~14.67), except that the integral now 
is finite and presented in the relativistically invariant form. Also, we have checked that the integration over $\phi$ can be effectively 
carried out even with the simplest trapezoid or Simpson formulas. Let us note that the two expressions for the Thomson amplitude 
[i.e., Eqs.~\eqref{thom4} and \eqref{thom11}] can be also used as a test when determining classical trajectories 
and evaluating the integral over $\phi$. We define next the phase of the complex Thomson amplitude
\begin{equation}
\Phi_{\mathrm{Th},\sigma}(\omega_{\bm{K}})=\arg(\mathcal{A}_{\mathrm{Th},\sigma})=\arg(\mathcal{A}^{\mathrm{inv}}_{\mathrm{Th},\sigma}) ,
\label{thom12}
\end{equation}
which is relativistically invariant but, in contrast to the Compton scattering, independent of spin degrees of freedom.

\begin{figure}
\includegraphics[width=8.5cm]{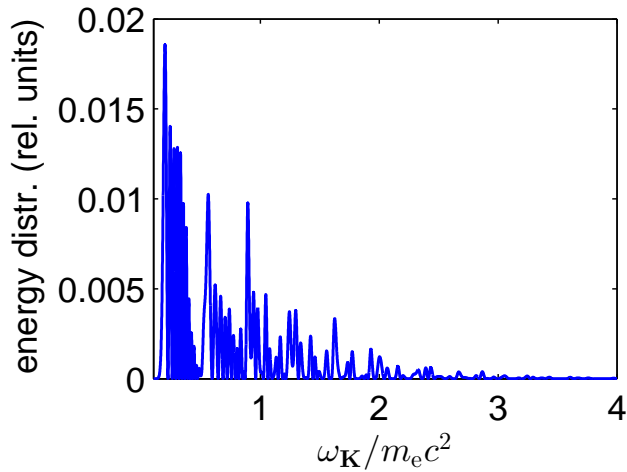}%
\caption{(Color online) The same as in Fig.~\ref{qspecf1xcomb20130914}, but for Thomson scattering. The derivative of the phase 
$\Phi_{\mathrm{Th},\sigma}(\omega_{\bm{K}})$ is not presented, as it is constant in the entire range of considered frequencies.
\label{cspecf1xcomb20130914}}
\end{figure}

In order to compare predictions of the classical and the quantum theories, let us study now the Thomson process for the same parameters 
as in Figs.~\ref{qspecf2comb20130914} and~\ref{qspecf1xcomb20130914}. For the parameters relevant to Fig.~\ref{qspecf2comb20130914}, 
the classical energy distribution is identical to the quantum energy distribution for the spin-conserved process. The difference 
between these two approaches shows up if we compare the corresponding phases, which for the classical theory linearly depends on the frequency of the generated radiation (meaning that its derivative 
is constant). The same happens for the parameters relevant to Fig.~\ref{qspecf1xcomb20130914}, for which the energy distribution is shown 
in Fig.~\ref{cspecf1xcomb20130914}. Let us remark that the energy distributions for the Compton (Fig.~\ref{qspecf1xcomb20130914}) and the 
Thomson (Fig.~\ref{cspecf1xcomb20130914}) processes, although not identical, are still
comparable in the sense that every peak or zero in these two distributions can unambiguously be related to each other~\cite{KKscale}. 
However, the corresponding phases depend on $\omega_{\bm{K}}$ differently. We would like to emphasize that the nonlinear dependence of the Compton 
phase on the frequency of emitted photons (contrary to the Thomson phase, which linearly depends on $\omega_{\bm{K}}$) is of the quantum origin. 
Such a qualitative difference between the classical and quantum results can be associated with a change of the electron final momentum in the 
Compton scattering, which introduces decoherence in the process. This fact, although in some cases unnoticed for the energy distribution of emitted 
photons, has far-reaching consequences for the temporal behavior of radiation generated by these two processes. This will be demonstrated in the next section.

Our main interest in this paper is nonlinear Compton scattering rather than its classical analogue, which is nonlinear Thomson scattering. 
The point being that the classical approach is an approximation of the complete quantum theory which takes into account the electron spin 
and the quantum recoil of electrons during the scattering. The complication being, that the quantum theory does now allow for 
as detailed description of the driving laser beam as the classical theory does. Also, it is more demanding computationally. For these reasons,
we investigate Thomson scattering for temporarily shaped laser pulses and for parameters for which both classical and quantum theories 
give either the same or different results. Our aim is to compare both theories in the context of short
pulse generation.

\section{Synthesis of short pulses}
\label{synthesis}

It is well-known that the energy distribution of generated radiation can be converted into the temporal power distribution. Currently, this is 
the standard technique used for the synthesis of attosecond pulses from the coherent combs of high-order harmonics. Here, let us consider the Thomson process and assume that the radiation is emitted in a given space direction, $\bm{n}_{\bm{K}}$. 
In this case, the temporal power distribution in the far radiation zone, which is remote from the scattering region by the distance $R$, is given by the formula (see, e.g. \cite{KKsuper})
\begin{equation}
\frac{\mathrm{d}^2P_{\mathrm{Th},\sigma}(\phi_{\mathrm{r}})}{\mathrm{d}^2\Omega_{\bm{K}}}=\frac{\alpha}{\pi}\bigl({\rm Re} \tilde{\mathcal{A}}^{(+)}_{\mathrm{Th},\sigma}(\phi_{\mathrm{r}})\bigr)^2 .
\label{ttt1}
\end{equation}
Here,
\begin{equation}
\tilde{\mathcal{A}}^{(+)}_{\mathrm{Th},\sigma}(\phi_{\mathrm{r}})=\int_0^{\infty}\mathrm{d}\omega \mathcal{A}_{\mathrm{Th},\sigma}(\omega)\mathrm{e}^{-\mathrm{i}\omega\phi_{\mathrm{r}}/\omega_0 }
\label{ttt2}
\end{equation}
is related to the electric field of the scattered radiation
\begin{equation}
\mathcal{E}_{\sigma}(\phi_{\mathrm{r}})=\frac{e}{4\pi\varepsilon_0cR}2{\rm Re} \tilde{\mathcal{A}}^{(+)}_{\mathrm{Th},\sigma}(\phi_{\mathrm{r}}),
\label{ttt3}
\end{equation}
where the symbol ``${\rm Re}$`` means the real value and $\sigma$ labels the polarization properties of emitted radiation. 
The quantity $\phi_{\mathrm{r}}$, which we call the retarded phase, is defined as
\begin{equation}
\phi_{\mathrm{r}}=\omega_0\Bigl(t-\frac{\bm{n}_{\bm{K}}\cdot \bm{r}}{c}\Bigr)=\omega_0\Bigl(t-\frac{R}{c}\Bigr) ,
\label{ttt4}
\end{equation}
with \textit{a~priori} an arbitrary real and positive $\omega_0$ that introduces the time-scale for the process. 
The retarded phase, for a given distance $R$ and $\omega_0$, determines the arrival-time of a light signal to the detector.

\begin{figure}
\includegraphics[width=8cm]{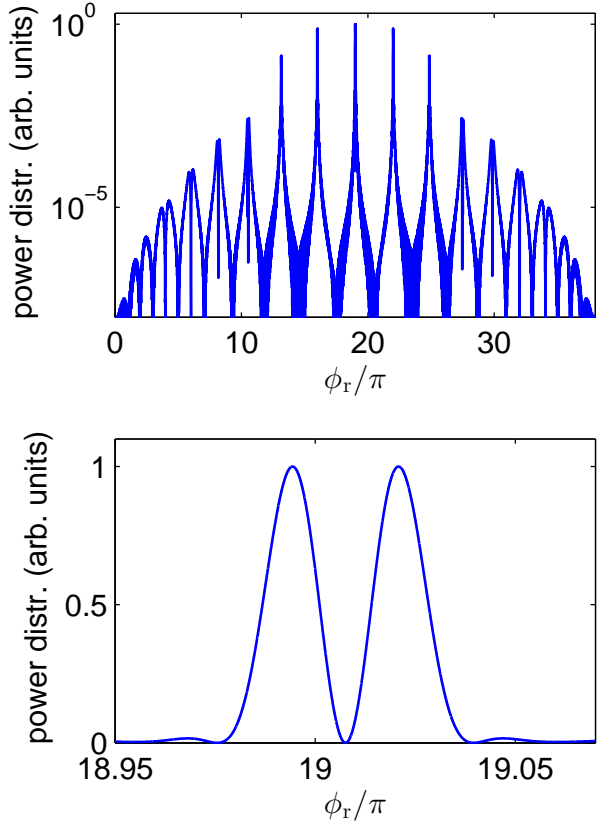}%
\caption{(Color online) Temporal power distribution, Eq.~\eqref{ttt1}, synthesized from the energy distribution for the Thomson process, 
when $\omega_0=\omega_{\mathrm{L}}$. The energy distribution is almost identical to that presented in Fig. \ref{qspecf2comb20130914} in the upper panel, 
except that the derivative of the Thomson phase is independent of the frequency. In the upper panel, the power distribution is presented 
in the logarithmic scale and it embraces the entire time-domain of the generated radiation. In the lower panel, the enlarged part of the central 
peak is shown in the linear scale. Since the temporal power distribution is proportional to the electric field of the emitted radiation squared, 
we see that the individual peaks form practically one-cycle pulses, which are well separated from each other. The distributions are scaled to their maximum value.
\label{cpuls16a02d201309019}}
\end{figure}
\begin{figure}
\includegraphics[width=8cm]{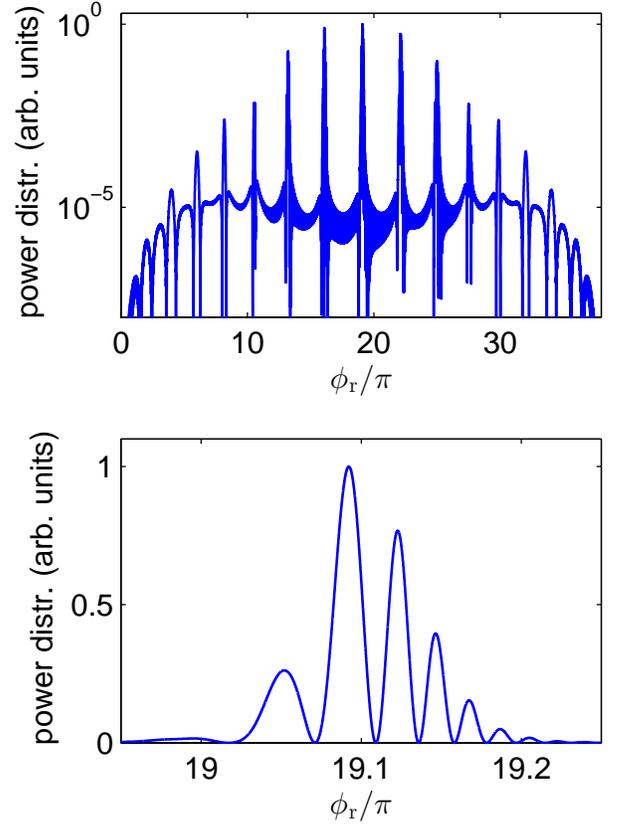}%
\caption{(Color online) The same as in Fig. \ref{cpuls16a02d201309019}, but for the Compton process such that $\lambda_{\mathrm{i}}\lambda_{\mathrm{f}}=1$. 
The nonlinear dependence of the Compton phase on the frequency $\omega_{\bm K}$ leads to a more complex structure of a given peak in the temporal power distribution. 
Instead of one-cycle pulses, as generated from Thomson scattering, now we observe many-cycle and chirped pulses of emitted radiation.
\label{qpuls16a02d201309019}}
\end{figure}

All the formulas presented in this section for the temporal power distributions also apply to the Compton process if in Eq.~\eqref{ttt2} 
the Thomson amplitude is replaced by the corresponding Compton amplitude. In this case, the power distribution depends not only on the polarization 
of emitted radiation but also on the spin degrees of freedom of the initial and final electrons.

For long laser pulses, the temporal power distribution could be a very rapidly oscillating function of time. For this reason, 
it is sometimes more convenient to consider the temporal power distribution averaged over such rapid oscillations, 
\begin{equation}
\frac{\mathrm{d}^2\langle P_{\mathrm{Th},\sigma}\rangle(\phi_{\mathrm{r}})}{\mathrm{d}^2\Omega_{\bm{K}}}=\frac{\alpha}{2\pi} |\tilde{\mathcal{A}}^{(+)}_{\mathrm{Th},\sigma}(\phi_{\mathrm{r}})|^2 ,
\label{ttt5}
\end{equation}
and similarly for the Compton process.

In Figs.~\ref{cpuls16a02d201309019} and~\ref{qpuls16a02d201309019}, we show the synthesis of the energy distribution from Thomson and Compton scattering, respectively. 
We see that temporal power distributions for both classical and quantum processes are qualitatively similar, as both consist of a sequence of very sharp peaks. 
However, individual peaks in each sequence are quite different, as presented in lower panels. For Thomson scattering (Fig.~\ref{cpuls16a02d201309019}), a peak consists practically of a single 
oscillation of the electric field. For Compton scattering (Fig.~\ref{qpuls16a02d201309019}), on the other hand, the structure of an individual peak is more complex. Namely, it represents a pulse 
of a few electric field oscillations with decreasing period. The origin of such a chirp is the nonlinear dependence of the Compton phase on $\omega_{\bm{K}}$. 
Note that this nonlinearity is the genuine quantum effect~\cite{KKscale}. Therefore, the chirp appearing in the generated radiation can be considered as a quantum signature 
in collisions of a non-chirped laser pulse with free electrons.

\begin{figure}
\includegraphics[width=8.5cm]{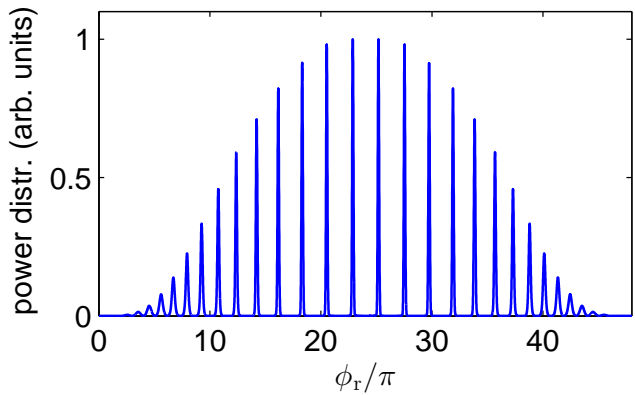}%
\caption{(Color online) Temporal power distribution, Eq.~\eqref{ttt1}, synthesized from the energy distribution of the Thomson process 
for the laser- and electron-beam parameters specified in Fig.~\ref{cspecf1xcomb20130914}, and for $\omega_0=\omega_{\mathrm{L}}$. Since the phase 
of the Thomson amplitude linearly depends on $\omega_{\bm{K}}$ we obtain the train of well-separated half-cycle pulses of emitted radiation.
The distribution is normalized to its maximum value.
\label{cpuls16x01d201309020}}
\end{figure}

\begin{figure}
\includegraphics[width=8.5cm]{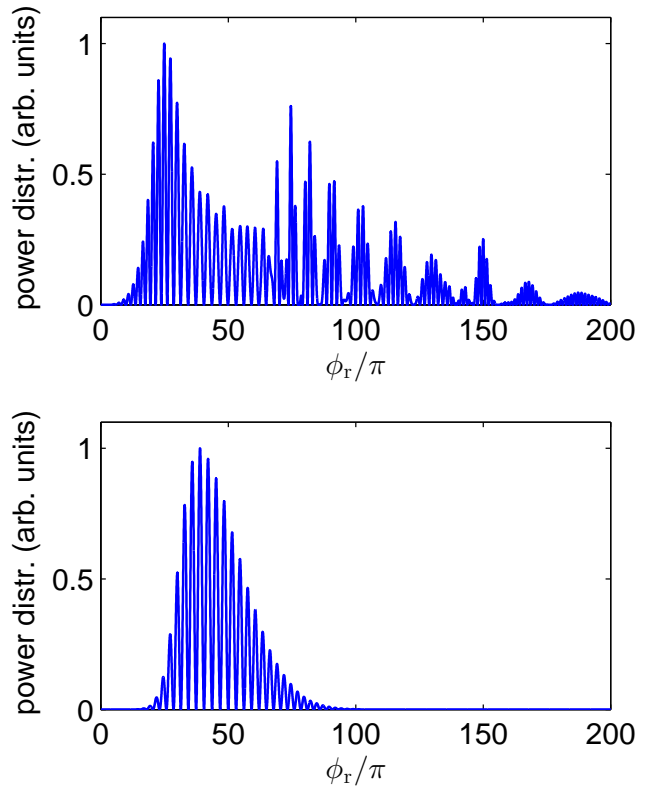}%
\caption{(Color online) Temporal power distribution synthesized from the energy distribution of the Compton process for the laser- and electron-beam parameters specified in
Fig.~\ref{qspecf1xcomb20130914}, and for $\omega_0=\omega_{\mathrm{L}}$ (upper panel). Since the phase of the Compton amplitude nonlinearly depends on $\omega_{\bm{K}}$ 
the emitted radiation does not form a train of short pulses. In the lower panel, the window-selected temporal power distribution [i.e., Eq.~\eqref{ttt7} with the window function \eqref{ttt9}] 
is presented for $\omega_{\mathrm{wmax}}=0.2m_{\mathrm{e}}c^2$. Although the window function selects only a 'regular' part from the energy distribution, 
nevertheless the corresponding temporal power distribution also does not exhibit a train of very short pulses. It rather represents 
a long pulse with many electric field oscillations. Both distributions are scaled to their maximum values.
\label{qpuls16x01d201309020}}
\end{figure}

Frequently, only a part of the spectrum of emitted radiation is used for the composition or 
detection of short laser pulses (see, for instance, the FROG technique~\cite{Trebino2000}). To account for this fact a window function 
(in the FROG it is called the gate function), $W(\omega)$, is introduced, which picks up a part of the frequency spectrum. 
The windowing of the emitted spectrum could also be related to the properties of detectors of radiation, that can be sensitive 
to frequencies from a particular range. In such a case, we define the window-selected amplitude
\begin{equation}
\tilde{\mathcal{A}}^{(+)}_{\mathrm{Th},\sigma}(\phi_{\mathrm{r}};W)=\int_0^{\infty}\mathrm{d}\omega W(\omega)\mathcal{A}_{\mathrm{Th},\sigma}(\omega)\mathrm{e}^{-\mathrm{i}\omega\phi_{\mathrm{r}}/\omega_0 },
\label{ttt6}
\end{equation}
so that the corresponding temporal power distributions are equal to
\begin{eqnarray}
\frac{\mathrm{d}^2P_{\mathrm{Th},\sigma}(\phi_{\mathrm{r}};W)}{\mathrm{d}^2\Omega_{\bm{K}}}&=&\frac{\alpha}{\pi}\bigl({\rm Re} \tilde{\mathcal{A}}^{(+)}_{\mathrm{Th},\sigma}(\phi_{\mathrm{r}};W)\bigr)^2,\label{ttt7}\\
\frac{\mathrm{d}^2\langle P_{\mathrm{Th},\sigma}\rangle(\phi_{\mathrm{r}};W)}{\mathrm{d}^2\Omega_{\bm{K}}}&=&\frac{\alpha}{2\pi} |\tilde{\mathcal{A}}^{(+)}_{\mathrm{Th},\sigma}(\phi_{\mathrm{r}};W)|^2 ,\label{ttt8}
\end{eqnarray}
and similarly for the Compton scattering.

Fig.~\ref{qpuls16a02d201309019} presents synthesized pulses in the case when the nonlinear dependence 
of the Compton phase on the frequency of scattered photons is small. To complement these results, we consider now the case 
of a strong dependence of the Compton phase on $\omega_{\bm{K}}$, i.e., for parameters specified in Fig.~\ref{qspecf1xcomb20130914}. Again, 
for these laser- and electron-beam parameters the derivative of the Thomson phase over $\omega_{\bm{K}}$ is constant, 
which leads to a very regular temporal power distribution of the generated radiation (see, Fig.~\ref{cpuls16x01d201309020}). 
Now, the individual peaks represent half-cycle pulses. We meet a completely different situation for the Compton process, 
for which the synthesis does not lead to a sequence of well-separated short pulses as it is in the case of the classical process.
Instead, we obtain a broad and irregular signal of emitted radiation, as shown in the upper panel of Fig.~\ref{qpuls16x01d201309020}. 
We want to emphasize that the reason for such a qualitative discrepancy between the classical and the quantum processes is 
the highly nonlinear dependence of the Compton phase on the frequency of emitted photons.

A question arises: Can the window-selecting help in producing trains of short pulses? To answer this question we consider the window function,
\begin{equation}
W(\omega)=\begin{cases}
0, & \omega < 0 \cr
\frac{1}{2}\bigl (1+\cos(\pi\omega/\omega_{\mathrm{wmax}})\bigr )  , & 0\leqslant\omega \leqslant \omega_{\mathrm{wmax}} \cr
0, & \omega > \omega_{\mathrm{wmax}}
\end{cases}
\label{ttt9}
\end{equation}
with $\omega_{\mathrm{wmax}}=0.2m_{\mathrm{e}}c^2$, such that it removes the irregular high-frequency part of the energy 
distribution shown in Fig.~\ref{qspecf1xcomb20130914}. The synthesized window-averaged temporal power distribution is presented in the lower panel 
of Fig.~\ref{qpuls16x01d201309020}. Indeed, we removed an irregular part of the power distribution for large retarded phases.
However, instead of a sequence of sharp spikes observed classically, we obtain the single pulse consisting of many regular 
oscillations of the electric field. The reason being that, for frequencies in the domain defined by the window function, 
the nonlinear terms in the Compton phase are still significant.

The great advantage of the classical approach is that calculations can be carried out quite easily, even for an arbitrary space 
and time dependent laser field. For this reason, the classical approach is extensively used in plasma physics and also in the context 
of ultra-short pulse generation \cite{Galkin2009,Chung2009,Lee2003,Lan2005,Kaplan2002,Liu2012}. Even though Thomson theory 
has some important shortcomings. For instance, it does not account for the spin of electrons, which for the high-frequency 
part of the spectrum starts to play a significant role \cite{KKpol,KKscale}, especially for very short and intense laser pulses. 
Another defect of the classical theory, which appears to be crucial for the extremely short pulse generation, is that it 
neglects the recoil of electrons during the emission of high-frequency photons \cite{Sarachik1970}. It has been noted that the electron recoil effects are small if
\begin{equation}
\omega_{\bm{K}} \ll \omega_{\mathrm{cut}}=c\frac{n\cdot p_{\mathrm{i}}}{n\cdot n_{\bm{K}}},
\label{rec1}
\end{equation}
independently of the laser field intensity, $I$, and also of the laser field carrier frequency, $\omega_{\mathrm{L}}$. On the other hand, 
it is well-known from the Fourier analysis that in order to generate the shorter radiation pulses the broader energy spectra have to 
be used for the pulse synthesis. There are two possibilities to increase the bandwidth of the energy distribution in Thomson or Compton 
processes. Namely, one can either increase the energy of electron beams or increase the intensity of the laser beam. Mostly, the second 
scenario is used \cite{Galkin2009,Lee2003,Lan2005,Kaplan2002,Liu2012}. The results presented in this section show that this scenario 
does not work for sufficiently intense laser pulses such that photons of frequencies comparable to $\omega_{\mathrm{cut}}$ are created 
with significant probabilities. Thus, conclusions drawn from the classical theory concerning the generation of extremely short 
radiation pulses, which are synthesized from frequencies close to the cut-off values, generally cannot be trusted.

\section{Frequency comb structures}
\label{combs}

\begin{figure}
\includegraphics[width=8.0cm]{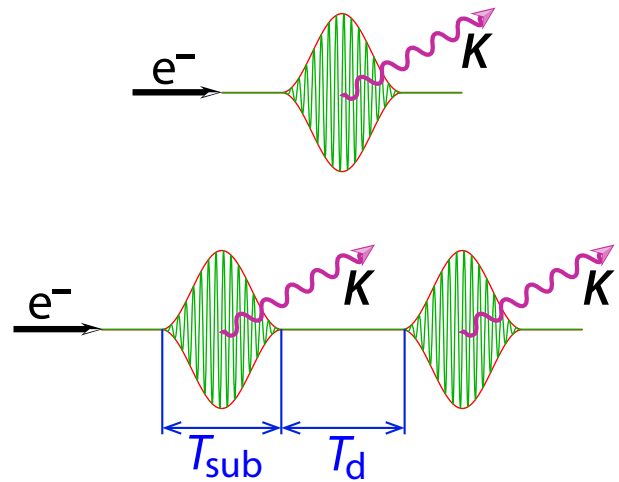}%
\caption{(Color online) Schematic diagram of Compton and Thomson scattering induced by a single (upper cartoon) and a double laser pulse with a time delay $T_{\mathrm{d}}$ (lower cartoon). 
The radiation emitted from each subpulse interfere leading to the formation of frequency combs in the energy distribution. 
The separation between peaks in the comb can be controlled by the time delay.
\label{yyoungmgtwofoto}}
\end{figure}

Discovery of the high-order harmonics in the interaction of laser pulses with atoms~\cite{exp1} and their subsequent theoretical 
analysis in terms of the three-step model~\cite{theo1} has stimulated a number of investigations. In particular, the coherent properties of the harmonics led Farkas and T\'oth~\cite{Farkas} to 
the idea of composing attosecond pulses from at least a part of the high-order harmonics comb. This is a routine method used currently
in attosecond physics~\cite{KrauszIvanov}. It was also shown that the three-step model is not the only mechanism 
responsible for the high-order harmonics generation and that such a comb of frequencies can be effectively generated by the channeling 
of initially unbounded electrons through crystal structures~\cite{FK1996}. In this case the emergence of multiple plateaus in the harmonics 
spectrum is due to resonance transitions between the laser-modified Floquet-Bloch states of electrons~\cite{FK1997} (very recently the Floguet-Bloch states have been detected experimentally \cite{FBexp}). 
A similar situation is met for the Thomson and Compton scattering, when the electron beam traverses the periodic structure 
of a laser beam (if approximated by a plane wave). This problem was extensively studied by Salamin and Faisal~\cite{SF1996,SF1997,SF2000} 
within classical theory. 

\begin{figure}
\includegraphics[width=7.5cm]{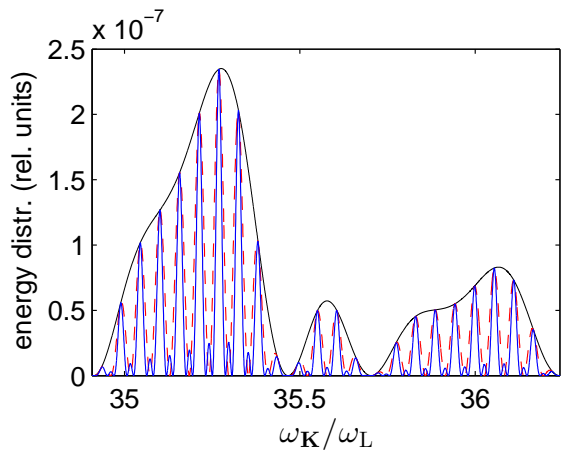}%
\caption{(Color online) Compton energy distribution, Eq.~\eqref{com2}, as a function of frequency $\omega_{\bm{K}}$ 
for $\lambda_{\mathrm{i}}\lambda_{\mathrm{f}}=1$. The laser beam, linearly polarized in the $x$-direction, propagates 
in the $z$-direction and collides with the electron beam in the head-on geometry. The distribution is calculated in the 
reference frame of electrons with the laser pulse parameters such that $\omega_{\mathrm{L}}=4.14\times 10^{-4}m_{\mathrm{e}}c^2$, 
$\mu=1$, $N_{\mathrm{osc}}=16$, $\chi=0$. The emitted radiation is calculated for $\theta_{\bm{K}}=0.3\pi$ and $\varphi_{\bm{K}}=0$. 
The thin black line (the envelope) corresponds to $N_{\mathrm{rep}}=1$, the thick dashed red line to $N_{\mathrm{rep}}=2$, the thick blue line 
to $N_{\mathrm{rep}}=3$, and the distributions are divided by $N_{\mathrm{rep}}^2$. The corresponding energy distribution for the Thomson 
process looks identical except that the classical one is blue-shifted by $0.2\omega_{\mathrm{L}}$.
\label{qcombprob20130921}}
\end{figure}

For short laser pulses the situation is different. Instead of sharp peaks, as the ones observed for long pulses, we observe broad coherent peak 
structures~\cite{KKsuper} extending to a few MeV. In our recent paper we demonstrated that, within such broad structures, it is possible to create 
coherent frequency combs for both the electromagnetic and the matter waves~\cite{KKcomb}. The idea is to use a modulated laser pulse, 
as illustrated in Fig.~\ref{yyoungmgtwofoto}. For instance, if we collide a sequence of two 
subpulses of duration $T_{\mathrm{sub}}$ each, and delayed by $T_{\mathrm{d}}$, with a nearly monochromatic electron beam (see, e.g., Ref.~\cite{KKcomb}), 
then the photons generated by each of these subpulses can interfere with each other. As a result, one might observe an interference pattern  in the energy distribution 
of emitted radiation. This is, of course, only the motivation and \textit{a priori} it is not obvious that the generated comb structures 
have similar coherent properties as the high-order harmonics combs. Only a numerical analysis of the Compton process can provide information 
about the phases of peaks within the comb and whether the Compton amplitudes can be synthesized to the finite and well-separated pulses;
this is indeed the case for the high-order harmonics combs. Note that the corresponding analysis of the classical Thomson process is insufficient.
First of all, because it is only an approximation of the quantum process. Secondly, as it follows from our discussion presented above, 
the phase properties of these two processes are in general different.

\begin{figure}
\includegraphics[width=8cm]{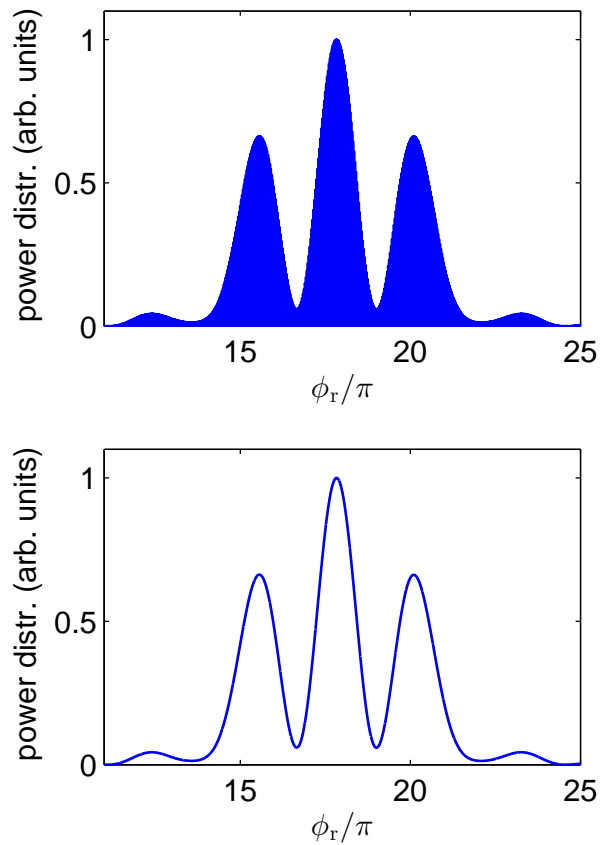}%
\caption{(Color online) Temporal power distribution (upper panel; Eq.~\eqref{ttt1} for the Compton scattering) for an unmodulated laser pulse, 
$N_{\mathrm{rep}}=1$. The remaining parameters are the same as in Fig.~\ref{qcombprob20130921}. The power distribution is synthesized 
from the energy distribution represented by the thin black line in Fig.~\ref{qcombprob20130921}. While this distribution shows very rapid oscillations,
in the lower panel it is averaged over these oscillations. Both distributions are normalized to their maximum values.
\label{qpuls16y01fuld201309020}}
\end{figure}

In Fig.~\ref{qcombprob20130921}, we present the Compton energy distribution for a particular range of frequencies of emitted photons 
and for the undelayed subpulses, $T_{\mathrm{d}}=0$. In this case, we obtain a broad structure which does not resemble 
the frequency comb. However, for $N_{\mathrm{rep}}>1$ the sharp peaks appear. They tend to become more narrow with increasing $N_{\mathrm{rep}}$, 
but they appear for the same frequencies independent of $N_{\mathrm{rep}}$. Moreover, the height of the individual peak scales as $N_{\mathrm{rep}}^2$, 
which already indicates the coherence of the generated comb. The numerical analysis of the phase of the Compton amplitude shows 
that at the peak frequencies phases are equal to 0 modulo $\pi$ \cite{KKcomb}. In addition, the derivative of the Compton phase with respect to $\omega_{\bm{K}}$
is almost constant (in the considered domain of $\omega_{\bm{K}}$). This proves that the separation between the consecutive peaks is nearly the same; 
hence, a coherent and equally spaced frequency comb is created.

\begin{figure}
\includegraphics[width=8cm]{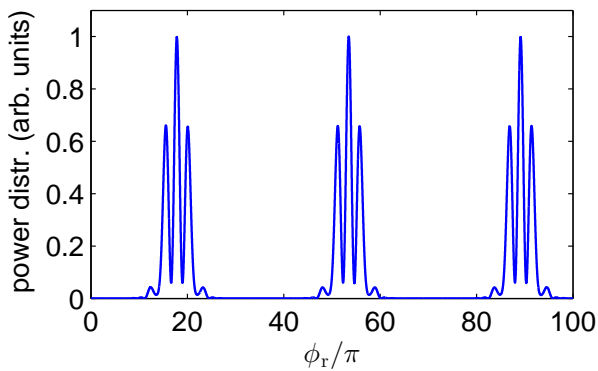}%
\caption{(Color online) Temporal power distribution averaged over the fast oscillations in the case of Compton scattering, for the same parameters 
as in Fig.~\ref{qpuls16y01fuld201309020} but for three subpulses, $N_{\mathrm{rep}}=3$. The synthesis of the corresponding energy distribution, 
represented in Fig.~\ref{qcombprob20130921} by the thick blue line, leads to a train of three identical pulses. The distribution is scaled to its maximum value.
\label{qpuls16y03avd201309020}}
\end{figure}

In the upper panel of Fig.~\ref{qpuls16y01fuld201309020}, we present the power distribution generated by a single pulse. As we see, the broad 
structure represented in Fig.~\ref{qcombprob20130921} by the envelope curve is converted into the rapidly oscillating and modulated pulse 
of radiation. The power distribution, averaged over these rapid oscillations, is shown in the lower panel of Fig.~\ref{qpuls16y01fuld201309020}. 
Note that the emitted pulse has a marginal chirp, which is the consequence of a very small nonlinearity in the dependence of the Compton phase 
on the frequency of created photons in the considered range of energies. Next, we synthesize the power distribution from the frequency 
comb generated by a sequence of $N_{\mathrm{rep}}$ pulses. As a result, we obtain nearly identical, well separated, and equally spaced in time 
$N_{\mathrm{rep}}$ copies of the same signal which was obtained for a single pulse. This is presented in Fig.~\ref{qpuls16y03avd201309020} for $N_{\mathrm{rep}}=3$. 
This proves the coherent properties of the frequency comb generated from nonlinear Compton (Thomson) process for this particular range of frequencies.

In Appendices \ref{diffthom} and \ref{diffcom} we derive the diffraction formulas for the Thomson and Compton amplitudes that prove 
the 'phase-matching' conditions for the peaks in the energy distributions at which the global phases change by $\pi$. We also show there 
that, although for classical theory this can happen for the equally spaced frequencies, for quantum theory this is not the case. 
The individual harmonics in frequency combs are approximately equally separated from each other only within finite frequency intervals, 
in which the nonlinear dependence of the Compton phase on the emitted photon frequency can be neglected.

\subsection{Combs for delayed subpulses}
\label{combsdelay}

The distance between peaks in the comb can be made smaller or, equivalently, the separation between the synthesized pulses of scattered radiation 
can be made larger, if subpulses are delayed with respect to each other. To illustrate this, we have to properly define the shape function 
(we denote it by $f_{\mathrm{d}}(\phi)$ for $0\leqslant \phi\leqslant 2\pi$ and 0 otherwise) for such a situation. Hence, we divide the duration 
of the pulse $T_{\mathrm{p}}$ into three pieces and, for simplicity, we assume that the outermost time intervals are equal. Such a situation 
is described by the following choice
\begin{equation}
f_{\mathrm{d}}(\phi)=
\begin{cases}
0, & 0\leqslant\phi\leqslant \xi\pi \cr
\bar{f}(\phi), & \xi\pi < \phi < 2\pi-\xi\pi \cr
0, & 2\pi-\xi\pi \leqslant \phi \leqslant 2\pi
\end{cases}
\label{del1}
\end{equation}
and 0 otherwise, where $0\leqslant \xi < 1$. This shape function is illustrated in the upper cartoon of Fig.~\ref{yyoungmgtwofoto}. If the pulse 
lasts for $T_{\mathrm{p}}$, then the time when it does not vanish is equal to $T_{\mathrm{sub}}=(1-\xi)T_{\mathrm{p}}$. For the function $\bar{f}(\phi)$ we choose
\begin{equation}
\bar{f}(\phi)=\sqrt{1-\xi\ }\ f\Bigl(\frac{\phi-\xi\pi}{1-\xi} \Bigr),
\label{del2}
\end{equation}
where
\begin{equation}
f^{\prime}(\phi)\propto\sin^2\Bigl(\frac{\phi}{2}\Bigr)\sin(N_{\mathrm{osc}}\phi+\chi),
\label{del3}
\end{equation}
as defined by Eq.~\eqref{las11} for $N_{\mathrm{rep}}=1$. Hence, the normalization condition, Eq.~\eqref{las6}, remains the same. Moreover, the central frequency 
of the laser field, $\omega_{\rm L}$, is related to the fundamental frequency, $\omega=2\pi/T_{\rm p}$, such that
\begin{equation}
\omega_{\mathrm{L}}=\frac{N_{\mathrm{osc}}\omega}{1-\xi}.
\label{del4}
\end{equation}
In order to form a sequence of $N_{\mathrm{rep}}$ subpulses, as illustrated in the lower cartoon of Fig.~\ref{yyoungmgtwofoto} for $N_{\mathrm{rep}}=2$, 
we have to repeat $N_{\mathrm{rep}}$ times the function~\eqref{del1}; this way we obtain subpulses with a time delay $T_{\mathrm{d}}=\xi T_{\mathrm{p}}$.
Then, we need to compress them back to the interval $[0,2\pi]$ remembering to divide the fundamental frequency and to multiply the laser central frequency by $N_{\mathrm{rep}}$.

\begin{figure}
\includegraphics[width=8cm]{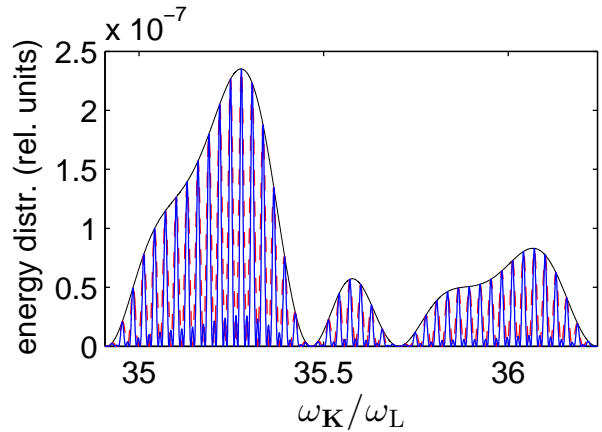}%
\caption{(Color online) The same as in Fig.~\ref{qcombprob20130921}, but for the delayed sequence of driving subpulses with $T_{\mathrm{sub}}=T_{\mathrm{d}}$ 
(cf., Fig.~\ref{yyoungmgtwofoto}). The delay between subpulses leads to a denser distribution of peaks in the frequency comb.
\label{qcombdelayprob20130923}}
\end{figure}

We remark that for a single pulse ($N_{\mathrm{rep}}=1$) the physical situation stays the same independently of which value for $\xi$ we choose.
The change of $\xi$ only means that we change the outermost time intervals, at which the electromagnetic field is 0. This means that all physical 
quantities including the energy distribution of emitted radiation (and, hence, the structure and the width of synthesized pulses) 
have to be the same. Only the time of creation of those pulses is shifted. This is a strong test for the correctness of the numerical analysis 
presented here. It has to be stressed, however, that for a nonzero $\xi$ the numerical calculations become much longer. The reason being that more Fourier components 
of the shape function have to be accounted for in order to properly approximate the vanishing parts of the driving pulse. The same applies to 
the sequence of driving subpulses.

In Fig.~\ref{qcombdelayprob20130923}, we present the energy distribution of generated Compton radiation for the same laser and electron beam 
parameters as in Fig.~\ref{qcombprob20130921}. This time, however, the driving subpulses are delayed by $T_{\mathrm{d}}=T_{\mathrm{sub}}$ 
(cf., Fig.~\ref{yyoungmgtwofoto}); in other words $\xi=0.5$. As we see, the results for $N_{\mathrm{rep}}=1$ are identical. On the other hand, the time delay 
between subpulses leads to a denser distribution of peaks in the frequency comb. Specifically, for the considered time delay the number of peaks 
doubles. The temporal power distribution also looks similar to the one shown in Fig.~\ref{qpuls16y03avd201309020}, except that the first pulse 
is delayed and the time distance to the next one is doubled. A very similar pattern is observed for Thomson scattering.

\section{Combs in laboratory frame}
\label{combslab}

The discussion above concerned Thomson and Compton processes when analyzed in the rest frame of electrons. This is a convenient reference frame for 
fundamental theoretical investigations, as most of geometrical degrees of freedom are eliminated and the analysis can focus mainly 
on dynamical aspects of these processes. From an experimental point of view, it is also not a serious limitation as the radiation generated during 
the collision of laser and electron beams interacts directly with the same electron beam. This was the case in the SLAC experiment~\cite{Bula}
in which electron-positron pairs had been generated by means of the Breit-Wheeler process (see, e.g.,~\cite{Reiss,Ritus,KKbw}). This takes place
in the cascade problems as well~\cite{Ruhl2013}. This means that properties of the generated radiation (such as chirping of the scattered radiation 
or the generation of frequency combs) can be detected indirectly by analyzing their consequences.

Apart from this, it is interesting to investigate properties of nonlinear Thomson and Compton scattering in the laboratory frame. It was shown~\cite{Galkin2009,Chung2009,Lee2003,Lan2005,Kaplan2002,Liu2012,KKsuper},
for instance, that in the laboratory frame the synthesis of generated radiation leads to zepto- or even yoctosecond pulses. This significantly extends 
the already well developed technique for attosecond pulse generation, which is based on the synthesis of coherent high-order harmonics combs \cite{Farkas}. 
The aim of this section is to investigate the possibility of direct detection of the frequency comb structures 
in the laboratory frame. In our analysis, we consider the Thomson scattering for the laser- and electron-beam parameters such that classical 
and quantum theories give similar results for the energy distribution of generated radiation. The reason for this limitation is that, from the numerical point of view, 
the classical calculation is much faster. A similar analysis for the Compton process is much more time-consuming and is going to be presented elsewhere in due course.

\begin{figure}
\includegraphics[width=8.5cm]{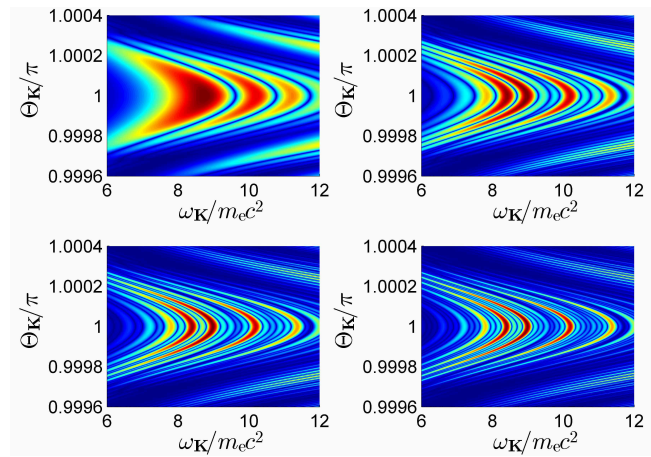}%
\caption{(Color online) Color mappings of the Thomson energy distribution produced in a head-on geometry of a laser beam and an electron beam. 
The electric field of a driving pulse, linearly polarized in the $x$-direction, is described by the shape function \eqref{las11} 
with $N_{\mathrm{osc}}=17$, and $N_{\mathrm{rep}}=1$ (upper left panel), $N_{\mathrm{rep}}=2$ (upper right panel), $N_{\mathrm{rep}}=3$ 
(lower left panel), $N_{\mathrm{rep}}=4$ (lower right panel). Its central frequency in the laboratory frame equals 
$\omega_{\mathrm{L}}=1.548\mathrm{eV}\approx 3\times 10^{-6}m_{\mathrm{e}}c^2$ and the averaged intensity is determined by $\mu^2=5/16$. 
Electrons move with momentum $\bm{p}_{\mathrm{i}}=1000m_{\mathrm{e}}c\,{\bm e}_z$ and the scattering process occurs in the plane $\Phi_{\bm{K}}=\pi/2$. 
The emitted radiation is linearly polarized in the $(xz)$-plane (or, equivalently, in the $(x'y')$-plane).
\label{surf17axrepalld20130906r600}}
\end{figure}

In order to obtain a significant signal of the emitted high-frequency radiation from Thomson or Compton scattering, when analyzed in the laboratory frame, 
the energy of the electron beam has to be sufficiently large. On the other hand, the central frequency of very intense laser pulses is much smaller 
than the rest mass of electrons. It follows from these two facts that the majority of Thomson (Compton) radiation is emitted into a very narrow cone. 
For the head-on collision of the laser and electron beams, this radiation is emitted mostly in the direction of the electron beam propagation. For this reason, 
it is better to parametrize the angular distribution of emitted radiation by a new set of angles. Let us change the Cartesian coordinates such that 
\begin{equation}
(x,y,z)\rightarrow (x',y',z')=(z,x,y),
\end{equation}
which is still a right-handed system of coordinates. Next, in the primed coordinates we introduce the polar, $0\leqslant\Phi_{\bm{K}}<\pi$, and azimuthal, 
$0\leqslant\Theta_{\bm{K}}\leqslant 2\pi$, angles. Hence, we find the following equations:
\begin{align}
\sin\Phi_{\bm{K}}\cos\Theta_{\bm{K}} = & \cos\theta_{\bm{K}}, \nonumber \\
\sin\Phi_{\bm{K}}\sin\Theta_{\bm{K}} = & \sin\theta_{\bm{K}}\cos\varphi_{\bm{K}}, \nonumber \\
\cos\Phi_{\bm{K}} = & \sin\theta_{\bm{K}}\sin\varphi_{\bm{K}},
\label{lab2}
\end{align}
which uniquely define a transformation between two pairs of angles. The scattering plane $(xz)$, which was defined before by two conditions, 
$\varphi_{\bm{K}}=0$ and $\varphi_{\bm{K}}=\pi$, now is defined by a single condition, $\Phi_{\bm{K}}=\pi/2$. The same parametrization was 
applied in our previous analysis of Compton scattering~\cite{KKcompton}. Note that now the measure of the solid angle is
\begin{equation}
\mathrm{d}^2\Omega_{\bm{K}}=\sin\Phi_{\bm{K}}\mathrm{d}\Phi_{\bm{K}}\mathrm{d}\Theta_{\bm{K}},
\label{lab3}
\end{equation}
where, for the considered head-on geometry, we can approximate $\sin\Phi_{\bm{K}}$ by 1 if integrating over a narrow angular cone.

\begin{figure}
\includegraphics[width=8.5cm]{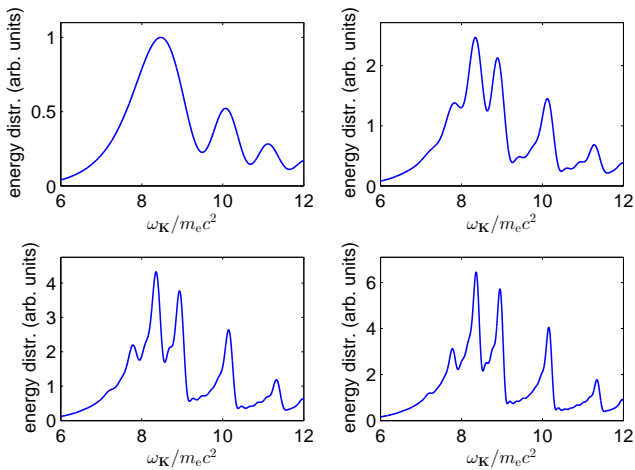}%
\caption{(Color online) The same as in Fig.~\ref{surf17axrepalld20130906r600}, but energy distributions are integrated over the 
angle $\Theta_{\bm{K}}$. They are normalized to the maximum value of the energy distribution for an unmodulated pulse (upper left panel). 
The comb peaks located at the same frequencies are clearly visible. Due to the integration over the polar angle, which introduces 
incoherence into the distribution, the maxima of the comb peaks do not scale as $N_{\mathrm{rep}}^2$. Nevertheless, the visibility 
of these peaks increase with increasing the number of subpulses.
\label{super17axrepalld20130906}}
\end{figure}

In Fig. \ref{surf17axrepalld20130906r600}, we present color mappings of the energy distribution of radiation generated in the scattering plane, 
$\Phi_{\bm{K}}=\pi/2$, for up to four repetitions ($N_{\mathrm{rep}}=1,2,3$ and 4) of a driving pulse without time delay, $T_{\mathrm{d}}=0$.
The results are for such frequencies $\omega_{\bm{K}}$ and angles $\Theta_{\bm{K}}$ for which most of the energy is emitted during the process. 
As expected, the energy is radiated in the very close vicinity of $\Theta_{\bm{K}}=\pi$. For a single pulse ($N_{\mathrm{rep}}=1$), we observe 
the formation of a broad hill for frequencies between 8 and 9$m_{\mathrm{e}}c^2$ (i.e., around 4MeV). The coherent properties of such structures 
(which in photonic physics are called the supercontinua~\cite{super2}) were considered elsewhere~\cite{KKsuper}. If, instead of a single 
pulse, we consider a sequence of such pulses then these broad structures are sliced into stripes and the coherent frequency comb is formed for a given angle
(see, the discussion in the previous section). These distributions integrated over the angle $\Theta_{\bm{K}}$,
\begin{equation}
\frac{\mathrm{d}^2E_{\mathrm{C}}}{\sin\Phi_{\bm{K}}\mathrm{d}\omega_{\bm{K}}\mathrm{d}\Phi_{\bm{K}}} =\int_0^{2\pi}\mathrm{d}\Theta_{\bm{K}}\frac{\mathrm{d}^3E_{\mathrm{C}}}{\mathrm{d}\omega_{\bm{K}}\mathrm{d}^2\Omega_{\bm{K}}} ,
\label{lab4}
\end{equation}
are presented in Fig.~\ref{super17axrepalld20130906}. We clearly see the formation of the comb peaks, whose positions stay the same 
for different number of subpulses. The maxima of these peaks increase with increasing $N_{\mathrm{rep}}$, although they do not scale like 
$N_{\mathrm{rep}}^2$. This is the signature of the incoherence caused by the integration over the angle, which also leads to the decrease 
of the visibility of the comb peaks in the integrated distribution. However, due to the large separation of these peaks and their comparable 
intensities, we are convinced that they could be detected experimentally. We remark that the survival of comb 
structures, even after integrating over angles, is due to significant collimation of the generated Thomson and Compton radiation, 
which happens for highly energetic electron beams.

\section{Conclusions}
\label{conclusions}

In this paper, we studied the nonlinear Thomson (classical theory) and the Compton (quantum theory) scattering of free electrons 
with temporarily finite laser pulses. We showed that, for the Compton spin-conserved process, the energy distribution of emitted 
photons can be well described by the classical Thomson theory provided that frequencies of generated photons are much smaller 
than the characteristic cut-off frequency. However, the phases of the corresponding classical and quantum amplitudes differ from each other. This 
results in different temporal power distributions for these two cases, although the corresponding energy distributions are nearly identical. 
Our analysis showed that, contrary to the classical theory, it is not always possible to synthesize short pulses from nonlinear Compton 
scattering. The point is that one has to choose the range of Compton photon frequencies in which nonlinear (or, equivalently, 
quantum) corrections to the Compton phase play a marginal role. This statement can be roughly quantified by the condition that
\begin{equation}
(\Delta\omega_{\bm{K}})^2\frac{\mathrm{d}^2}{\mathrm{d}^2\omega_{\bm{K}}}\Phi_{\mathrm{C},\sigma}(\omega_{\bm{K}},\lambda_{\mathrm{i}},\lambda_{\mathrm{f}}) \ll 1,
\label{con1}
\end{equation}
for $\lambda_{\mathrm{i}}\lambda_{\mathrm{f}}=1$, where $\Delta\omega_{\bm{K}}$ is the frequency bandwidth used for the synthesis 
of generated pulses of radiation or, in other words, the nonlinear corrections to the Compton phase within the frequency bandwidth 
are very small. The condition above is violated, for instance, for parameters specified in Fig.~\ref{qspecf2comb20130914}, although 
the frequencies are much smaller than $\omega_{\mathrm{cut}}$ and the classical and quantum energy distributions are almost identical.

In addition, we investigated a possibility of generating coherent frequency combs from Thomson and Compton scattering in the presence of a sequence of short subpulses. 
This was motivated by the celebrated high-order harmonic generation and by the resulting synthesis of attosecond pulses out of the frequency spectrum of those harmonics combs. We showed that the separation of peaks 
in the Compton-based (Thomson-based) frequency comb can be controlled by a time delay of subpulses. Note that such a control is not possible for the high-order harmonics, 
for which the distance between peaks is not smaller than the central frequency of the driving pulse, $\omega_{\mathrm{L}}$. The possible 
generation of a sequence of short pulses has also been investigated. In this context, as follows from our previous considerations~\cite{KKsuper}, 
the nonlinear Thomson and Compton processes provide the unique mechanism for the generation of zepto- or even yoctosecond pulses. Moreover, 
by analyzing nonlinear Thomson scattering in the laboratory frame, we presented a clear signature of the frequency comb in the angle-integrated 
energy distribution of emitted radiation, which could be detected experimentally.

We studied here the generation of frequency-comb structures for the ideal situation when all subpulses are identical. 
Such a situation can be well-modeled by composing laser pulses from a few monochromatic ones. In fact, the laser pulse shapes 
considered in this paper are composed from three monochromatic components with appropriately chosen amplitudes, and from only 
two of such components one can build the sequence of identical subpulses for $N_{\mathrm{osc}}=2$. This fact raises the question: 
How sensitive is the formation of frequency combs if we change relative phases of these monochromatic components? This and similar 
problems are currently investigated and are going to be presented in due course.

\section*{Acknowledgements}

This work is supported by the Polish National Science Center (NCN) under Grant No. 2012/05/B/ST2/02547. 

\appendix
\section{Triads of unit vectors}
\label{triads}

The aim of this appendix is to settle the convention for the polarization vectors for both the laser pulse and the radiation emitted 
during Thomson or Compton scattering. Let us define three normalized and mutually orthogonal real vectors, $\bm{a}_j$, $j=1,2,3$, such that
\begin{equation}
\bm{a}_1=\begin{pmatrix}\cos\theta\cos\varphi \cr \cos\theta\sin\varphi \cr -\sin\theta \end{pmatrix} \!\! ,
\bm{a}_2=\begin{pmatrix}-\sin\varphi \cr \cos\varphi \cr 0 \end{pmatrix} \!\! ,
\bm{a}_3=\begin{pmatrix}\sin\theta\cos\varphi \cr \sin\theta\sin\varphi \cr  \cos\theta \end{pmatrix} \!\! ,
\label{app1}
\end{equation}
where $\theta$ and $\varphi$ are the polar and azimuthal angles in an arbitrary chosen reference frame.
These vectors constitute a right-handed basis of vectors, since
\begin{equation}
\bm{a}_i=\varepsilon_{ijl}\bm{a}_j\times\bm{a}_l,
\label{app2}
\end{equation}
where $\varepsilon_{ijl}$ is the antisymmetric tensor such that $\varepsilon_{123}=1$.
Moreover, an arbitrary vector $\bm{V}$ can be decomposed as
\begin{equation}
\bm{V}=\bm{a}_1 (\bm{a}_1\cdot\bm{V})+\bm{a}_2 (\bm{a}_2\cdot\bm{V})+\bm{a}_3 (\bm{a}_3\cdot\bm{V}).
\label{app3}
\end{equation}

Usually, we shall assume that, if the radiation propagates in the $\bm{a}_3$ direction, then two real vectors, 
$\bm{a}_1$ and $\bm{a}_2$, describe two linear polarizations of radiation. In order to account for 
elliptic polarizations, we should consider two linear combinations,
\begin{eqnarray}
\bm{a}_{\delta,1}&=&\cos\delta \, \bm{a}_1 +\mathrm{i}\sin\delta\,  \bm{a}_2,\label{app4}\\
\bm{a}_{\delta,2}&=&\mathrm{i}\sin\delta \, \bm{a}_1 +\cos\delta\,  \bm{a}_2,\label{app5}
\end{eqnarray} 
such that the orthogonality condition reads $\bm{a}_{\delta,j}\cdot \bm{a}_{\delta,l}^*=\delta_{jl}$. In this case, 
the right-handed condition, Eq. \eqref{app2}, remains valid and, for an arbitrary vector $\bm{V}$, the following decomposition is fulfilled:
\begin{equation}
\bm{V}=\bm{a}_{\delta,1} (\bm{a}_{\delta,1}^*\cdot\bm{V})+\bm{a}_{\delta,2} (\bm{a}_{\delta,2}^*\cdot\bm{V})+\bm{a}_3 (\bm{a}_3\cdot\bm{V}).
\label{app6}
\end{equation}
In particular, for $\delta=\pi/4$ the vectors $\bm{a}_{\delta,1}$ and $\bm{a}_{\delta,2}$ correspond to the right-handed and left-handed circular polarizations.

Note that the choice of vectors $\bm{a}_1$ and $\bm{a}_2$ in Eq.~\eqref{app1} is not unique. We can use this freedom to define 
another set of vectors which determines the polarization properties of a beam of photons propagating in different directions. 
If, for instance, we have a polarizer which does not transmit radiation polarized perpendicular to the unit vector $\bm{N}_{\mathrm{pol}}$,
then it is sometimes more convenient to introduce a triad of vectors $(\bm{a}_{\|},\bm{a}_{\bot},\bm{n})$ such that
\begin{equation}
\bm{a}_{\|}=v_1\bm{a}_1+v_2\bm{a}_2,\ \bm{a}_{\bot}=-v_2\bm{a}_1+v_1\bm{a}_2,\ \bm{a}_{\|}\times\bm{a}_{\bot}=\bm{n},
\label{app7}
\end{equation}
where
\begin{equation}
v_i=\frac{\bm{a}_i\cdot\bm{N}_{\mathrm{pol}}}{\sqrt{(\bm{a}_1\cdot\bm{N}_{\mathrm{pol}})^2+(\bm{a}_2\cdot\bm{N}_{\mathrm{pol}})^2}}, \ i=1,2.
\label{app8}
\end{equation}

\section{Diffraction and global phase for Thomson scattering}
\label{diffthom}

We derive here the diffraction formula for classical Thomson scattering that resembles very much the diffraction grating formula for angular distributions. For this purpose let us consider an arbitrary pulse defined by two shape functions $f_{0j}(\phi)$ ($j=1,2$ for two linear polarizations of the laser field) such that they vanish outside the interval $[0,2\pi/N_{\mathrm{rep}}]$ together with their first derivatives, and for $N_{\mathrm{rep}}=1,2,\dots$. If we define now the shape functions $f_j(\phi)$ in Eq.~\eqref{las1} such that
\begin{equation}
f_j(\phi)=\begin{cases} f_{0j}(\phi), & \phi\in [0,2\pi/N_{\mathrm{rep}}], \cr
                        0, & \mathrm{otherwise},
				  \end{cases}
\label{bbi1}
\end{equation}
then the Thomson formula, Eq.~\eqref{thom4}, defines the energy distribution for a single pulse.
Since the acceleration of electrons for $\phi>2\pi/N_{\mathrm{rep}}$ vanishes, therefore the upper 
limit of the integration over the phase $\phi$ can be shrunk to $2\pi/N_{\mathrm{rep}}$. On the other hand, the shape functions
\begin{equation}
f_j(\phi+2\pi (L-1)/N_{\mathrm{rep}})=f_{0j}(\phi), \, \mathrm{for}\, L=1,2,\dots,N_{\mathrm{rep}},
\label{bbi2}
\end{equation}
define the pulse consisting of $N_{\mathrm{rep}}$ copies of the same subpulse. In this situation,
\begin{align}
 \mathcal{A}_{\mathrm{Th},\sigma}(\omega_{\bm{K}})=&\frac{1}{2\pi}\int_0^{2\pi}\mathrm{d}\phi\Upsilon_{\sigma}(\phi)
 \mathrm{e}^{\mathrm{i}\omega_{\bm{K}}\ell(\phi)/c} \nonumber \\
 =&\frac{1}{2\pi}\sum_{L=1}^{N_{\mathrm{rep}}}\int_0^{2\pi/N_{\mathrm{rep}}}\mathrm{d}\phi
 \Upsilon_{\sigma}\Bigl(\phi+2\pi\frac{L-1}{N_{\mathrm{rep}}}\Bigr) \nonumber \\
 &\qquad\times\exp\Bigl[\mathrm{i}\frac{\omega_{\bm{K}}}{c}\ell\Bigl(\phi+2\pi\frac{L-1}{N_{\mathrm{rep}}}\Bigr)\Bigr] .
\label{bi1}
\end{align}
For $0\leqslant\phi\leqslant 2\pi/N_{\mathrm{rep}}$,
\begin{equation}
 \Upsilon_{\sigma}\Bigl(\phi+2\pi\frac{L-1}{N_{\mathrm{rep}}}\Bigr)=\Upsilon_{\sigma}(\phi)
\label{bi2} 
\end{equation}
and
\begin{equation}
 \ell\Bigl(\phi+2\pi\frac{L-1}{N_{\mathrm{rep}}}\Bigr)=(L-1)\ell\Bigl(\frac{2\pi}{N_{\mathrm{rep}}}\Bigr)+\ell(\phi).
\label{bi3} 
\end{equation}
Hence, after some algebraic manipulations, we arrive at the \textit{diffraction formula} for the Thomson amplitude,
\begin{align}
 \mathcal{A}_{\mathrm{Th},\sigma}(\omega_{\bm{K}})=&\exp\Bigl[\mathrm{i}\frac{\omega_{\bm{K}}}{2c}(N_{\mathrm{rep}}-1)
 \ell\Bigl(\frac{2\pi}{N_{\mathrm{rep}}}\Bigr)\Bigr] \nonumber \\
 &\times\frac{\sin\Bigl[\frac{\omega_{\bm{K}}N_{\mathrm{rep}}}{2c}\ell\Bigl(\frac{2\pi}{N_{\mathrm{rep}}}\Bigr)\Bigr]}
 {\sin\Bigl[\frac{\omega_{\bm{K}}}{2c}\ell\Bigl(\frac{2\pi}{N_{\mathrm{rep}}}\Bigr) \Bigr]}\mathcal{A}^{(1)}_{\mathrm{Th},\sigma}(\omega_{\bm{K}})
\label{bi4}
\end{align}
where (see, Eq.~\eqref{thom3} with the comments below \eqref{bbi1})
\begin{equation}
 \mathcal{A}^{(1)}_{\mathrm{Th},\sigma}(\omega_{\bm{K}})=\frac{1}{2\pi}\int_0^{2\pi/N_{\mathrm{rep}}}\mathrm{d}\phi\Upsilon_{\sigma}(\phi)
 \mathrm{e}^{\mathrm{i}\omega_{\bm{K}}\ell(\phi)/c}
\label{bi5} 
\end{equation}
is the Thomson amplitude for the single subpulse.

For particular frequencies $\omega_{\bm{K},L}$ that fulfill the condition
\begin{equation}
 \frac{\omega_{\bm{K},L}}{c}\ell\Bigl(\frac{2\pi}{N_{\mathrm{rep}}}\Bigr)=2\pi L,\quad L=1,2,\dots\, ,
\label{bi6}
\end{equation}
we have the diffraction enhancement of the energy distribution generated by Thomson scattering 
(similar to the diffraction grating pattern for the angular distribution), as
\begin{equation}
 |\mathcal{A}_{\mathrm{Th},\sigma}(\omega_{\bm{K},L})|^2=N_{\mathrm{rep}}^2 |\mathcal{A}^{(1)}_{\mathrm{Th},\sigma}(\omega_{\bm{K},L})|^2.
 \label{bi7}
\end{equation}
Moreover, for $N_{\mathrm{rep}}>1$, the Thomson amplitude vanish for $\omega_{\bm{K}}$ such that
\begin{equation}
 \frac{\omega_{\bm{K}}N_{\mathrm{rep}}}{2c}\ell\Bigl(\frac{2\pi}{N_{\mathrm{rep}}}\Bigr)=\pi L,\quad L=1,\dots,N_{\mathrm{rep}}-1,
 \label{bi8}
\end{equation}
and, for $N_{\mathrm{rep}}>2$, it has minor maxima if
\begin{equation}
 \frac{\omega_{\bm{K}}N_{\mathrm{rep}}}{2c}\ell\Bigl(\frac{2\pi}{N_{\mathrm{rep}}}\Bigr)=\pi L
 +\frac{\pi}{2},\quad L=1,\dots,N_{\mathrm{rep}}-2.
 \label{bi9}
\end{equation}
This pattern is exactly observed in our numerical analysis and is very well-known for the angular distribution of radiation passing through the diffraction grating.

\begin{figure}
\includegraphics[width=8cm]{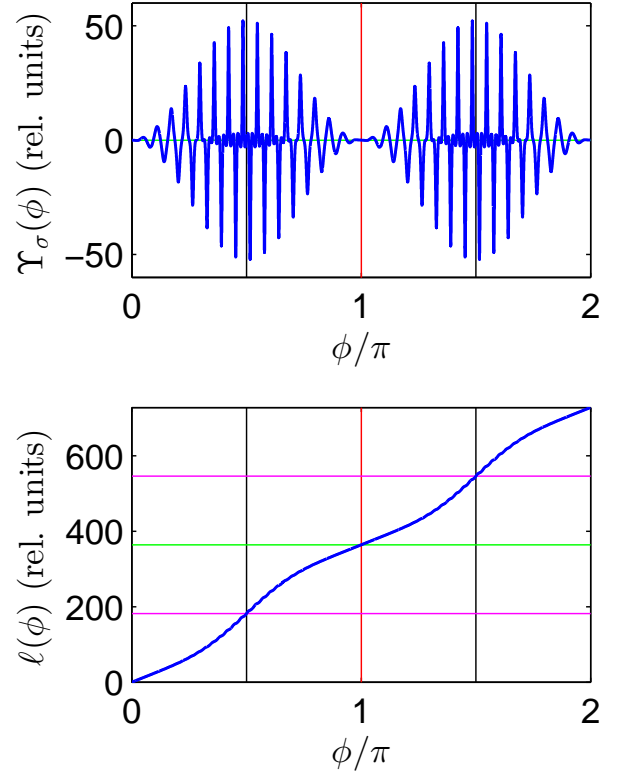}%
\caption{(Color online) Functions $\Upsilon_{\sigma}(\phi)$ and $\ell(\phi)$ for the Thomson amplitude.
The parameters are the same as in Fig.~\ref{qspecf1xcomb20130914} except that $N_{\mathrm{rep}}=2$. 
These functions, for the considered laser pulse shapes, satisfy the symmetry conditions \eqref{bi11} and \eqref{bi12}. 
We draw horizontal and vertical lines to emphasize the important symmetries of these functions. 
\label{spec16x01d20140207}}
\end{figure}

The global phase of Thomson amplitude equals
\begin{align}
\arg \mathcal{A}_{\mathrm{Th},\sigma}(\omega_{\bm{K}})=&(N_{\mathrm{rep}}-1)\Bigl[\pi+
\frac{\omega_{\bm{K}}}{2c} \ell\Bigl(\frac{2\pi}{N_{\mathrm{rep}}}\Bigr)\Bigr] \nonumber \\
+&\arg \mathcal{A}^{(1)}_{\mathrm{Th},\sigma}(\omega_{\bm{K}}),
\label{bi10}
\end{align}
and the determination of the phase for a single subpulse for a general form of the shape functions and arbitrary polarizations of emitted radiation can 
be done only numerically. However, for special types of pulses considered in this paper the analytical formula for this phase can be provided. Indeed, by inspecting Fig.~\ref{spec16x01d20140207}, together with the comments made below Eq.~\eqref{bi1}, one can notice the following symmetry properties, valid for $\phi\in [0,\pi/N_{\mathrm{rep}}]$,
\begin{equation}
\Upsilon_{\sigma}(\phi)=-\Upsilon_{\sigma}(2\pi/N_{\mathrm{rep}}-\phi)
\label{bi11}
\end{equation}
and
\begin{equation}
\ell(\phi)+\ell(2\pi/N_{\mathrm{rep}}-\phi)=2\ell(\pi/N_{\mathrm{rep}}).
\label{bi12}
\end{equation}
These relations allow us to write down the Thomson amplitude $\mathcal{A}^{(1)}_{\mathrm{Th},\sigma}(\omega_{\bm{K}})$ as follows:
\begin{align}
\mathcal{A}^{(1)}_{\mathrm{Th},\sigma}&(\omega_{\bm{K}})=\frac{1}{\pi}\exp\Bigl[\mathrm{i}
\Bigl(\frac{\pi}{2}+\frac{\ell(\pi/N_{\mathrm{rep}})}{c}\omega_{\bm{K}}\Bigr)\Bigr]
\nonumber \\
\times & \int_0^{\pi/N_{\mathrm{rep}}}\mathrm{d}\phi\, \Upsilon_{\sigma}(\phi)\sin\Bigl[\frac{\omega_{\bm{K}}}{c}\bigl(\ell(\phi)-\ell(\pi/N_{\mathrm{rep}})\bigr) \Bigr],
\label{phase3}
\end{align}
and, since $\ell(2\pi/N_{\mathrm{rep}})=2\ell(\pi/N_{\mathrm{rep}})$, we finally arrive at the global phase for Thomson amplitude,
\begin{equation}
\Phi_{\mathrm{Th},\sigma}(\omega_{\bm{K}})=\Bigl(N_{\mathrm{rep}}\mp\frac{1}{2}\Bigr)\pi+
N_{\mathrm{rep}}\frac{\omega_{\bm{K}}}{c} \ell\Bigl(\frac{\pi}{N_{\mathrm{rep}}}\Bigr),
\label{bi13}
\end{equation}
where ``$-$`` is if the integral in \eqref{phase3} is positive, and ``$+$`` if negative. Therefore, we see that for laser pulses considered 
in this paper the global phase is a linear function of the frequency of emitted radiation and, moreover, for the peak frequencies $\omega_{\bm{K},L}$, Eq.~\eqref{bi6}, we obtain,
\begin{equation}
\Phi_{\mathrm{Th},\sigma}(\omega_{\bm{K},L})=\Bigl(N_{\mathrm{rep}}\mp\frac{1}{2}\Bigr)\pi+N_{\mathrm{rep}}L\pi.
\label{bi14}
\end{equation}
Hence, up to the same constant term, the phase is 0 modulo $\pi$, which proves the coherent properties of the Thomson combs. 
Moreover, the peak frequencies $\omega_{\bm{K},L}$ are equally separated from each other, which is not the case for Compton scattering. 

We remark that, in order to derive the diffraction formula \eqref{bi4}, one has to assume that for each individual subpulse
all necessary conditions imposed on a laser pulse have to be preserved; namely, the electromagnetic field strength and vector potential 
in the beginning and at the end of a subpulse has to vanish. Otherwise, the symmetry relations \eqref{bi2} and \eqref{bi3} would 
not be satisfied. The same applies to the quantum case, as it follows from analysis presented in Appendix~\ref{diffcom}. 
This, in particular, excludes the case of a plane wave as for the single oscillation these conditions are not satisfied.

\section{Diffraction and global phase for Compton scattering}
\label{diffcom}

A similar analysis as in Appendix~\ref{diffthom}, can be also carried out for Compton scattering. Since in this case the formulas 
are much longer, we first introduce simplified notations. In this Appendix the integers $j,j'=1,2$ denote two linear 
polarizations of the laser pulse, and we apply the Einstein summation convention. Let us also define the following abbreviations:
\begin{equation}
\mu_{\mathrm{i}}=\mu\frac{m_{\mathrm{e}}c}{2p_{\mathrm{i}}\cdot k}, \quad \mu_{\mathrm{f}}=\mu\frac{m_{\mathrm{e}}c}{2p_{\mathrm{f}}\cdot k},
\label{ci1}
\end{equation}
\begin{equation}
S^{(+)}_p(x)=p\cdot x+\int_0^{k\cdot x}\Bigl[\frac{eA(\phi)\cdot p}{k\cdot p}
-\frac{e^2A^2(\phi)}{2p\cdot k} \Bigr]\mathrm{d}\phi,
\label{ci2}
\end{equation}
and the four-vector
\begin{equation}
Q=p_{\mathrm{i}}-p_{\mathrm{f}}-K.
\label{ci3}
\end{equation}
Then the probability amplitude for Compton scattering can be written as \cite{KKcompton}
\begin{equation}
\mathcal{A}(e^-_{p_{\mathrm{i}}\lambda_{\mathrm{i}}}\rightarrow e^-_{p_{\mathrm{f}}\lambda_{\mathrm{f}}}
+\gamma_{\bm{K}\sigma})=\mathrm{i}\sqrt{\frac{2\pi\alpha c(m_{\mathrm{e}}c^2)^2}{E_{\bm{p}_{\mathrm{i}}}E_{\bm{p}_{\mathrm{f}}}\omega_{\bm{K}}V^3}}\, \mathcal{A},
\label{ci4}
\end{equation}
where $V$ is the quantization volume and
\begin{equation}
\mathcal{A}=\int\mathrm{d}^4x\mathrm{e}^{-\mathrm{i}(S^{(+)}_{p_{\mathrm{i}}}(x)
-S^{(+)}_{p_{\mathrm{f}}}(x)-K\cdot x)}
\bar{u}^{(+)}_{\bm{p}_{\mathrm{i}}\lambda_{\mathrm{i}}}\hat{\cal{C}}(k\cdot x)
u^{(+)}_{\bm{p}_{\mathrm{f}}\lambda_{\mathrm{f}}},
\label{ci5}
\end{equation}
with the $4\times 4$ matrix function
\begin{align}
\hat{\cal{C}}(k\cdot x)=&\slashed{\varepsilon}_{\bm{K}\sigma}-\mu_{\mathrm{i}}f_j(k\cdot x)\slashed{\varepsilon}_{\bm{K}\sigma}\slashed{k}\slashed{\varepsilon}_j-\mu_{\mathrm{f}}f_j(k\cdot x)\slashed{\varepsilon}_j\slashed{k}\slashed{\varepsilon}_{\bm{K}\sigma}
\nonumber \\
+&\mu_{\mathrm{i}}\mu_{\mathrm{f}}f_j(k\cdot x)f_{j'}(k\cdot x)\slashed{\varepsilon}_j\slashed{k}\slashed{\varepsilon}_{\bm{K}\sigma}\slashed{k}\slashed{\varepsilon}_{j'}.
\label{ci6}
\end{align}
For finite laser pulses this expression, although finite, is not convenient for numerical and analytical analysis. 
Therefore, we apply the transformation defined in Appendix B in Ref.~\cite{KKcompton}, and originally introduced 
by Boca and Florescu in Ref.~\cite{puls2}. This transformation leads to the change of $\hat{\cal{C}}(k\cdot x)$,
\begin{align}
\hat{\cal{C}}(k\cdot x)=&\bigl(\tilde{a}_jf_j(k\cdot x)+\tilde{b}[f_1^2(k\cdot x)+f_2^2(k\cdot x)]\bigr)\slashed{\varepsilon}_{\bm{K}\sigma}
\nonumber \\
-&\mu_{\mathrm{i}}f_j(k\cdot x)\slashed{\varepsilon}_{\bm{K}\sigma}\slashed{k}\slashed{\varepsilon}_j-\mu_{\mathrm{f}}f_j(k\cdot x)\slashed{\varepsilon}_j\slashed{k}\slashed{\varepsilon}_{\bm{K}\sigma}
\nonumber \\
+&\mu_{\mathrm{i}}\mu_{\mathrm{f}}f_j(k\cdot x)f_{j'}(k\cdot x)\slashed{\varepsilon}_j\slashed{k}\slashed{\varepsilon}_{\bm{K}\sigma}\slashed{k}\slashed{\varepsilon}_{j'}.
\label{cci6}
\end{align}
Here,
\begin{equation}
\tilde{a}_j=2\frac{k^0}{Q^0}(\mu_{\mathrm{i}}p_{\mathrm{i}}\cdot\varepsilon_j-\mu_{\mathrm{f}}p_{\mathrm{f}}\cdot\varepsilon_j),
\label{cci6a}
\end{equation}
and
\begin{equation}
\tilde{b}=-\frac{k^0}{Q^0}\mu m_{\mathrm{e}}c(\mu_{\mathrm{i}}-\mu_{\mathrm{f}}).
\label{cci6b}
\end{equation}
Now, accounting for the laser pulse-dressed electron momentum, Eq.~\eqref{com0}, 
we introduce the following decomposition (this is in fact the definition of $G(k\cdot x)$),
\begin{equation}
S^{(+)}_{p_{\mathrm{i}}}(x)-S^{(+)}_{p_{\mathrm{f}}}(x)-K\cdot x=\bar{Q}\cdot x+G(k\cdot x),
\label{ci9}
\end{equation}
where
\begin{equation}
\bar{Q}=\bar{p}_{\mathrm{i}}-\bar{p}_{\mathrm{f}}-K.
\label{ci10}
\end{equation}
The purpose of this decomposition is such that the functions $G(\phi)$ and $\hat{\cal{C}}(\phi)$ for the laser pulse consisting of $N_{\mathrm{rep}}$ copies of identical subpulses satisfy, for $\phi\in [0,2\pi/N_{\mathrm{rep}}]$ and $L=1,\dots,N_{\mathrm{rep}}-1$, the symmetry conditions
\begin{equation}
G(\phi+2\pi L/N_{\mathrm{rep}})=G(\phi),
\label{ci11}
\end{equation}
and
\begin{equation}
\hat{\cal{C}}(\phi+2\pi L/N_{\mathrm{rep}})=\hat{\cal{C}}(\phi),
\label{ci12}
\end{equation}
similar to Eq.~\eqref{bi2} for Thomson scattering. Further, for an arbitrary four-vector $a$, we define the light-cone variables ($\bm{n}$ is the propagation direction of the laser beam)
\begin{equation}
a^{\|}=\bm{n}\cdot\bm{a},\, a^-=a^0-a^{\|},\, a^+=\frac{a^0+a^{\|}}{2},\, \bm{a}^{\bot}=\bm{a}-a^{\|}\bm{n}.
\label{ci13}
\end{equation}
Since ($x^-=k\cdot x/k^0=\phi/k^0$)
\begin{equation}
\bar{Q}\cdot x=(\bar{Q}^+/k^0)\phi+\bar{Q}^-x^{-} -\bar{\bm{Q}}^{\bot}\cdot\bm{x}^{\bot},
\label{ci14}
\end{equation}
and
\begin{equation}
\mathrm{d}^4x=\frac{1}{k^0}\mathrm{d}\phi\mathrm{d}x^+\mathrm{d}^2x^{\bot},
\label{ci15}
\end{equation}
we rewrite the Compton amplitude \eqref{ci5} as
\begin{align}
\mathcal{A}=&(2\pi)^3\delta(Q^-)\delta^{(2)}(\bm{Q}^{\bot})\frac{1}{k^0} \nonumber \\
\times & \int_0^{2\pi}\mathrm{d}\phi \,\mathrm{e}^{-\mathrm{i}(\bar{Q}^+/k^0)\phi}\,
\mathrm{e}^{-\mathrm{i}G(\phi)}\bar{u}^{(+)}_{\bm{p}_{\mathrm{i}}\lambda_{\mathrm{i}}}\hat{\cal{C}}(\phi)
u^{(+)}_{\bm{p}_{\mathrm{f}}\lambda_{\mathrm{f}}}
\label{ci16}
\end{align}
Applying now the decomposition \eqref{bi1} to the integral over $\phi$ and accounting for the symmetries \eqref{ci11} and \eqref{ci12}, 
we arrive finally at the \textit{diffraction formula} for the Compton amplitude,
\begin{equation}
\mathcal{A}=\exp\Bigl(-\mathrm{i}\pi\frac{\bar{Q}^+(N_{\mathrm{rep}}-1)}{k^0N_{\mathrm{rep}}}\Bigr)
\frac{\sin(\pi\bar{Q}^+/k^0)}{\sin(\pi\bar{Q}^+/k^0N_{\mathrm{rep}})}\mathcal{A}^{(1)},
\label{ci17}
\end{equation}
where $\mathcal{A}^{(1)}$ is the Compton amplitude for a single pulse. For frequencies of emitted photons, $\omega_{\bm{K},L}$ with integer $L$, that satisfy the condition
\begin{equation}
\pi\bar{Q}^+/k^0N_{\mathrm{rep}}=-\pi L,
\label{ci18}
\end{equation}
we have the coherent enhancement of the Compton amplitude, which leads to the quadratic, $N_{\mathrm{rep}}^2$, enhancement of probability distributions. 
However, contrary to the Thomson case, these frequencies are not \textit{exactly} equally separated from each other on the whole interval of allowed 
frequencies, i.e. $[0,\omega_{\mathrm{cut}}]$. When $\omega_{\bm{K}}$ approaches the cut-off value $\omega_{\mathrm{cut}}$ the spectrum of $\omega_{\bm{K},L}$ 
becomes increasingly denser. This means that one can get the frequency comb for Compton scattering with approximately equally spaced peak frequencies, only over some limited frequency intervals.

Since for a single subpulse (see, discussion in Sec.~\ref{compton})
\begin{equation}
\arg\mathrm{A}^{(1)}=-\pi\frac{\bar{Q}^+}{k^0N_{\mathrm{rep}}}+\Phi_{\mathrm{C},\sigma}^{\mathrm{dyn}}(\omega_{\bm{K}},\lambda_{\mathrm{i}},\lambda_{\mathrm{f}}),
\label{ci19}
\end{equation}
where $\Phi_{\mathrm{C},\sigma}^{\mathrm{dyn}}(\omega_{\bm{K}},\lambda_{\mathrm{i}},\lambda_{\mathrm{f}})$ is the dynamic phase of a single subpulse, 
therefore the global phase of the Compton amplitude equals
\begin{equation}
\arg\mathcal{A}=\Phi_{\mathrm{C},\sigma}(\omega_{\bm{K}},\lambda_{\mathrm{i}},\lambda_{\mathrm{f}})=
-\pi\frac{\bar{Q}^+}{k^0}+\Phi_{\mathrm{C},\sigma}^{\mathrm{dyn}}(\omega_{\bm{K}},\lambda_{\mathrm{i}},\lambda_{\mathrm{f}}).
\label{ci20}
\end{equation}

For arbitrary laser pulses and polarizations of emitted photons the dynamic phase can only be calculated numerically. 
We have checked numerically that for laser pulses considered in this paper the dynamic phase is independent of $\omega_{\bm{K}}$. 
Hence, for the peak frequencies $\omega_{\bm{K},L}$, the global phase, 
\begin{equation}
\Phi_{\mathrm{C},\sigma}(\omega_{\bm{K},L},\lambda_{\mathrm{i}},\lambda_{\mathrm{f}})=
\pi LN_{\mathrm{rep}}+\Phi_{\mathrm{C},\sigma}^{\mathrm{dyn}}(\omega_{\bm{K},L},\lambda_{\mathrm{i}},\lambda_{\mathrm{f}}),
\label{ci21}
\end{equation}
is the same modulo $\pi$. This does not mean, however, that the Compton frequency comb, contrary to the Thomson one, is perfectly 
coherent. This time, the distance between the peaks change a little bit, due to the recoil of electrons during the emission of photons. 
For the low-frequency part of the frequency spectrum these effects are rather small, but for the high-frequency part they become significant.

\end{document}